\documentclass[journal,twoside,web]{ieeecolor}
\usepackage{generic}
\usepackage{cite}
\usepackage{lmodern}
\usepackage{courier}
\usepackage{amsmath,amssymb,amsfonts}
\usepackage{graphicx}
\usepackage{subfigure}
\usepackage[utf8]{inputenc}
\usepackage[swedish]{babel}
\usepackage{lmodern}
\usepackage{mathtools}
\usepackage{cuted}
\usepackage[justification=centering]{caption}
\usepackage{xcolor}
\usepackage{xfrac}
\usepackage{algorithm,algorithmicx}
\usepackage{algpseudocode}

\DeclareMathOperator*{\prox}{\mathbf{prox}}
\newcommand{\R}{\mathbb{R}}

\newcommand{\vct}[1]{\boldsymbol{#1}}
\newcommand{\mtx}[1]{\boldsymbol{#1}}

\newcommand{\Trsp}{\mathrm{T}}

\newcommand{\set}[1]{\mathcal{#1}}

\newcommand{\linop}[1]{\mathcal{#1}}	

\newcommand{\vc}{\vct{c}}

\newcommand{\ve}{\vct{e}}

\newcommand{\vh}{\vct{h}}

\newcommand{\vl}{\vct{l}}

\newcommand{\vp}{\vct{p}}
\newcommand{\vq}{\vct{q}}

\newcommand{\vu}{\vct{u}}
\newcommand{\vv}{\vct{v}}

\newcommand{\vy}{\vct{y}}

\newcommand{\vsigma}{\vct{\sigma}}

\newcommand{\vepsilon}{\vct{\epsilon}}

\newcommand{\vnu}{\vct{\nu}}

\newcommand{\vzeta}{\vct{\zeta}}
\newcommand{\vxi}{\vct{\xi}}

\newcommand{\vzero}{\vct{0}}

\newcommand{\mA}{\mtx{A}}
\newcommand{\mB}{\mtx{B}}

\newcommand{\mE}{\mtx{E}}

\newcommand{\mH}{\mtx{H}}
\newcommand{\mI}{\mtx{I}}

\newcommand{\mP}{\mtx{P}}
\newcommand{\mQ}{\mtx{Q}}

\newcommand{\mT}{\mtx{T}}
\newcommand{\mU}{\mtx{U}}
\newcommand{\mV}{\mtx{V}}

\newcommand{\mX}{\mtx{X}}
\newcommand{\mY}{\mtx{Y}}
\newcommand{\mZ}{\mtx{Z}}
\newcommand{\mDelta}{\mtx{\Delta}}

\newcommand{\mGamma}{\mtx{\Gamma}}

\newcommand{\mPi}{\mtx{\Pi}}

\newcommand{\mId}{{\bf I}}

\newcommand{\mzero}{{\bf 0}}

\newcommand{\loG}{\linop{G}}

\newcommand{\loP}{\linop{P}}

\newcommand{\setS}{\set{S}}

\newcommand{\setW}{\set{W}}

\def\n{\natural}

\usepackage[T1]{fontenc}
\usepackage{fancyhdr}
\usepackage[babel=true]{microtype}
\usepackage{textcomp}
\def\BibTeX{{\rm B\kern-.05em{\sc i\kern-.025em b}\kern-.08em
    T\kern-.1667em\lower.7ex\hbox{E}\kern-.125emX}}
\fancypagestyle{alim}{\fancyhf{}\fancyfoot[L]{\footnotesize \textit{This artcle has been accepted at IEEE Transactions on Biomedical Engineering (TBME), 2025.}}}

\begin{document}

\title{Stationary and Sparse Denoising Approach for Corticomuscular Causality Estimation}
\author{Farwa Abbas,  Verity McClelland, Zoran Cvetkovic, and Wei Dai
\thanks{Farwa Abbas (f.abbas20@imperial.ac.uk) and Wei Dai (wei.dai1@imperial.ac.uk) are currently with the Electrical Engineering Department, Imperial College London, UK.}
\thanks{Verity McClelland (verity.mcclelland@kcl.ac.uk) is with King's College London, UK. She also continues clinical work as an honorary Consultant in Clinical Neurophysiology at Evelina London Children's Hospital and Great Ormond Street Hospital, London, UK.}
\thanks{Zoran Cvetkovic (zoran.cvetkovic@kcl.ac.uk) is with the Electrical Engineering Department, King's College London. \\ {\fontfamily{ptm}\selectfont Copyright \textcopyright\ 2021 IEEE. Personal use of this material is permitted. However, permission to use this material for any other purposes must be obtained from the IEEE by sending an email to pubs-permissions@ieee.org.}}
}
\maketitle

\begin{abstract}
\textit{Objective:} Cortico-muscular communication patterns are instrumental in understanding movement control. Estimating significant causal relationships between motor cortex electroencephalogram (EEG) and surface electromyogram (sEMG) from concurrently active muscles presents a formidable challenge since the relevant processes underlying muscle control are typically weak in comparison to  measurement noise and background activities. \textit{Methodology:} In this paper, a novel framework is proposed to simultaneously estimate the order of the autoregressive model of cortico-muscular interactions along with the parameters while enforcing stationarity condition in a convex program to ensure global optimality. The proposed method is further extended to a non-convex program to account for the presence of measurement noise in the recorded signals by introducing a wavelet sparsity assumption on the excitation noise in the model.
\textit{Results:} The proposed methodology is validated using both simulated data and neurophysiological signals. In case of simulated data, the performance of the proposed methods has been compared with the benchmark approaches in terms of order identification, computational efficiency, and goodness of fit in relation to various noise levels. In case of physiological signals our proposed methods are compared against the state-of-the-art approaches in terms of the ability to detect Granger causality. \textit{Significance:} The proposed methods are shown to be effective in handling stationarity and measurement noise assumptions, revealing significant causal interactions from brain to muscles and vice versa. 
\end{abstract}


\section{Introduction}
\label{sec:introduction}

\IEEEPARstart{I}{nference} of causal interactions involved in movement control from   non-invasive physiological signals such as EEG (electroencephalography) and sEMG (surface electromyography) has been 
immensely important in neurology \cite{delezie2018endocrine}, prosthetics \cite{chowdhury2019eeg}, cognitive neuroscience \cite{fuchs2009embodied}, biomechanics \cite{chiel2009brain}, robotics \cite{qiao2021survey}, and  related disciplines. In the context of movement disorders, understanding how the brain and muscles interact is fundamental in diagnosing and treating  disorders such as Parkinson's disease, multiple sclerosis, and dystonia. These conditions often involve neurological dysfunction or disruptions in the communication between the brain and muscles. 
Numerous studies have explored analysis of concurrent recordings of EEG from the primary motor cortex and sEMG signals from limb muscles to extract information about cortico-muscular coupling 
\cite{mcclelland2012modulation,du2021dictionary,mcclelland2020abnormal,witham2011contributions,zandvoort2019human}.

\thispagestyle{alim}

{A}{utoregressive} models have commonly been employed to capture the relationships between different processes and discern directional causality \cite{hsiao1982autoregressive},\cite{lutkepohl2005new},\cite{goebel2003investigating}. These models reveal causal relationships in diverse applications in computer science such as microservice architectures \cite{ahmad2024smart, ahmad2024towards}, aiding in threat detection \cite{goel2024machine, ahmad2023review} and attack tracing \cite{ahmad2024survey}, while large language models \cite{ahmad2025future, chopra2024chatnvd, haque2022think} enhance this by analyzing logs and predicting vulnerabilities \cite{abdulsatar2024towards, jayalath2024microservice}. The growing complexity of distributed systems has made such causal analysis increasingly vital for understanding system behavior and maintaining reliability.
The notion of causality, as applied to autoregressive models, was originally introduced by Granger \cite{granger2001investigating} where the fundamental concept is to quantify the degree to which knowledge about one process reduces uncertainty associated with another. However,  there are numerous restrictive assumptions \cite{shojaie2022granger} that must be met for the autoregressive model to be considered a suitable framework for identifying Granger causal relationships.  

Firstly, an appropriate model order, that quantifies the relevant process history for prediction, is assumed to be known a priori for Granger causality analysis \cite{shojaie2022granger}. Numerous  techniques have been put forward for determining the appropriate order of the autoregressive model. The optimal model order should be  sufficiently small to guarantee that predictions depend primarily on direct observations rather than interim estimations, yet large enough to encompass the inherent trend within the underlying process. When Granger causality is estimated using a small model order, it tends to limit the frequency resolution of physiological signals, making it challenging to distinguish effectively between different frequency ranges \cite{florin2016parkinson}. Traditionally, model identification is performed through visual assessment of the autocorrelation function or by minimizing loss functions like negative log-likelihood or least squares loss. Popular methods like the Akaike Information Criterion (AIC) \cite{akaike1969fitting} and Bayesian Information Criterion (BIC) \cite{schwarz1978estimating} have followed this approach. However, these methods may struggle to accurately infer the true data distribution \cite{shmueli2010explain} and can potentially overfit in practical scenarios \cite{cruz2006good}. More recently, a method for simultaneous model order identification and parameter estimation for ARMA (AutoRegressive Moving Average) models is proposed \cite{liu2020fitting}. This approach employs a sparsity-based regularization technique to estimate the order and leverages a Block Coordinate Descent based method 
for parameter estimation. This regularization has proven effective in determining the model order by progressively curtailing the irrelevant history of the process in a hierarchical manner, however, the method does not guarantee convergence to a global optimal point.

Estimating Granger causality also necessitates the assumption of stationarity as a fundamental prerequisite for the autoregressive process \cite{shojaie2022granger, abbas2024dlgc}, since it helps ensure that the historical patterns observed in the data are likely to persist.
In most prior research, the process is either assumed to be stationary \cite{songsiri2013sparse} or methods such as detrending \cite{watson1986univariate} and differencing \cite{lutkepohl2005new} are employed as heuristics to induce stationarity within the model. A recent advancement in this domain \cite{liu2020fitting}, introduced a method that computes parameter estimates and later it finds Euclidean projections onto a stationary subspace. As described in \cite{liu2020fitting}, projections onto such a stationary subspace may not be unique due to the potential non-convexities of the subspace. 
Nevertheless, the authors in \cite{liu2020fitting} resort to Proximal Block Coordinate Descent (BCD) to implement a projection-based approach.

In contrast to that, we proposed to use of the stability condition to enforce stationarity. It is well established that a Vector AutoRegressive VAR(p) model can be written as a first-order VAR(1) model by making use of the companion-form matrix. If all eigenvalues of the companion-form matrix are less than unity in absolute value, then the VAR model is stable and, hence, covariance stationary. However, the converse is not true i.e. an unstable VAR process can be stationary \cite{graupe1980convergence}.


Another key assumption made by autoregressive models is that the recorded variables are perfectly observed without any measurement errors \cite{shojaie2022granger}.These errors originate from external sources unrelated to the electrical activity of brain (or muscles), such as electrical interference from nearby devices or imperfections in the recording equipment. When dealing with physiological signals such as EEG and sEMG, the recorded signals are corrupted by measurement noise that is not necessarily negligible \cite{patriota2010vector}. Moreover, due to the specific nature of the structure inherent in autoregressive signals, using the ordinary time-series denoising methods \cite{alyasseri2021eeg, abbas2024robust} might not be as effective. Denoising the autoregressive processes is a challenging ill-posed problem that has been traditionally looked at from an errors-in-variables perspective \cite{patriota2010vector}, \cite{guo2023structured}. However, for mathematical tractability, the existing research imposes additional assumptions on the input signal, dynamic system, and the measurement noise terms that may not be applicable in real-world scenarios \cite{soderstrom2003errors}.  Another approach is to introduce structural assumptions on the excitation noise signal in the autoregressive model to obtain robust estimates \cite{christmas2010robust}. Excitation noise (or \textit{innovations}) typically represents any residual variability in the signal that is not accounted for by the autoregressive component, mainly arising from other physiological processes. This includes muscle activity, eye movements, or cardiac rhythms, which can introduce unwanted signals into the recording signal. Additionally, whilst the EEG recording  includes brain activity relating to cortex-muscle communication, it will also contain information relating to other processes, which in this context represent physiological \textit{noise}.  Introducing structural assumptions about excitation noise helps distinguish between different types of noise within the signal. This separation allows for the incorporation of theoretical or empirical knowledge about the system being modeled, making it easier to identify and filter out the noise from the underlying signal.

{ It has been noted that the classical Granger Causality approach using autoregressive models is prone to bias \cite{stokes2017study, antonacci2020information, faes2017interpretability , antonacci2021estimation}, and an alternative approach of using state-space models may be more robust as these have been shown to better handle some of these biases and constraints. However, the classical models remain a powerful and practical option in many applications.  In this study we use autoregressive models because they are computationally efficient, relatively easy to implement, and well-suited for capturing short-term linear temporal relationships. }
 
 Therefore, in this paper we propose two  methods to overcome the above issues related to model order estimation,  stationarity and  noise in observed processes.  Building upon the sparsity-based approach introduced in \cite{liu2020fitting}, we propose its advancement to determine the order of the model \textit{while} estimating parameters instead of projecting afterwards.
  We then propose to combine order identification with  a regularization that promotes stable estimates by effectively enforcing stationarity on the estimate -- we refer to this approach as 
  Stationary and Sparse Alternating Direction Method of Multipliers (SS-ADMM). To the best of our knowledge, this represents the first instance where the stationary condition is explicitly enforced within a convex program to ensure global optimality. The proposed method is further refined to take into account the effect of measurement noise on recorded signals. To that end, structural assumptions are employed to regularize the problem, enabling efficient solution via a non-convex ADMM solver. The resulting algorithm is termed as Stationary and Sparse Denoising ADMM (SSD-ADMM).  The goal of this research is to accurately model the recorded data to eventually improve the sensitivity of the inferred Granger causality estimates even in the presence of measurement noise.
The paper thus  makes the following contributions:
\begin{enumerate}
\item A unified approach is developed, that concurrently determines the model order and parameter estimates within a convex optimization framework, ensuring stability and global optimality. We refer to this approach as Stationary and Sparse Alternating Direction Method of Multipliers -- SS-ADMM.

\item The proximal operator for regularization based on stationarity condition is proposed that allows for efficient computation. 
\item A modification of  SS-ADMM is developed to effectively handle measurement noise in the recorded signals by employing a non-convex ADMM solver. 
\end{enumerate}

The paper is organized as follows.  The problem is formulated in Section II. In Section III, we present our SS-ADMM method, and in Section IV, we develop our SSD-ADMM method that takes into account measurement noise. Evaluation results are presented in Section V. Our findings are discussed in Section VI and conclusions are drawn in Section VII. A preliminary version of this work has been reported in \cite{abbas2023ss}, which is extended here by employing a non-convex ADMM solver to handle measurement noise.

\section{Problem Formulation}

In  autoregressive modeling, the sEMG signal can be modelled as a linear combination of lagged values of relevant EEG control activity, previous muscle activity, and excitation noise from a random innovation process. The EEG signal can also be expressed in a similar way with sensory feedback proportionate to sEMG activity sent back to the cortex. 

Let $y(t)$ be the EEG signal and $x(t)$ be the corresponding sEMG signal at time $t$. They can be represented as:
\small{
\begin{align}
\label{eq1}y(t) = \sum_{k=1}^{m} a_{yy}^{(k)}y(t-k)+\sum_{k=1}^{m} a_{yx}^{(k)}x(t-k)+\epsilon_y(t), \\
\label{eq2}x(t) = \sum_{k=1}^{m} a_{xy}^{(k)}y(t-k)+\sum_{k=1}^{m} a_{xx}^{(k)}x(t-k)+\epsilon_x(t),
\end{align}}\normalsize
where $\epsilon_y(t)$ and $\epsilon_x(t)$ are time-dependent excitation noise signals and $m$ indicates the dependency on past instants or order of the autoregressive model. 
The objective is to fit an autoregressive model to estimate parameters $a_{yy}^{(k)},a_{yx}^{(k)},a_{xy}^{(k)},a_{xx}^{(k)}$ $\forall k=1,\cdots,m$ where $m$ is known a priori. However, in practice, determining the  order of the autoregressive model is a non-trivial problem.

The system of equations in (\ref{eq1}-\ref{eq2}) can be compactly represented as follows:
\small{\begin{align}
\label{eq3}\begin{bmatrix}
y(t) \\ x(t) 
\end{bmatrix} &= \begin{bmatrix}a_{yy}^{(1)}\cdots a_{yy}^{(m)}&a_{yx}^{(1)}\cdots a_{yx}^{(m)}\\a_{xy}^{(1)}\cdots a_{xy}^{(m)}&a_{xx}^{(1)}\cdots a_{xx}^{(m)}
\end{bmatrix}\begin{bmatrix}y(t-1)\\\vdots \\y(t-m)\\x(t-1)\\\vdots \\x(t-m)
\end{bmatrix} +\begin{bmatrix}\epsilon_y(t) \\ \epsilon_x(t)\end{bmatrix}, \\
\label{eq4}\mY &= \mA \mH + \mE,
\end{align}} \normalsize

\noindent
where the matrices in Equation (\ref{eq3}) are assigned to corresponding variables in Equation (\ref{eq4}) for $t=m+1,...,m+T$. Hence, $\mY \in \R^{2 \times T}$, $\mA \in \R^{2 \times 2m}$, $\mH \in \R^{2m \times T}$ and $\mE \in \R^{2 \times T}$.  The model in Equation (\ref{eq1}) is the termed as \textit{unrestricted} model. On the other hand, a \textit{restricted} model would be the one that does not involve any cross-coupling between $x(t)$ and $y(t)$ i.e. $a_{xy}^{(i)}=a_{yx}^{(i)}=0$, $i=1, \cdots, m$. 

Mathematically, the \textit{restricted} model amounts to the one below:
\small{\setlength{\arraycolsep}{2pt}
  \renewcommand{\arraystretch}{0.8}\begin{align}
\label{eq5}\begin{bmatrix}y(t) \\ x(t) \end{bmatrix} & = \begin{bmatrix}a_{yy}^{(1) ^\prime}\cdots a_{yy}^{(m)^\prime}&  0\cdots\cdots\cdot\cdot0 \\0\cdots\cdots\cdot\cdot 0 &a_{xx}^{(1)^\prime}\cdots a_{xx}^{(m)^\prime}
\end{bmatrix}\begin{bmatrix}y(t-1) \\\vdots \\y(t-m) \\ x(t-1)\\ \vdots \\ x(t-m)
\end{bmatrix} +\begin{bmatrix}\epsilon^\prime_y(t) \\ \epsilon^\prime_x(t)\end{bmatrix}, \\
\label{eq6}\mY &= {\mA}^\prime {\mH}+ {\mE}^\prime,
\end{align}}\normalsize
where the matrices in Equation (\ref{eq5}) are assigned to corresponding variables in Equation (\ref{eq6}) and $\mA^\prime \in \R^{2 \times 2m}$ and $\mE^\prime \in \R^{2 \times T}$ are the coefficients and noise matrices for restricted model, respectively. Granger causality can be tested using an F-test \cite{shojaie2022granger} comparing the two models as: 
\begin{align}
\text{GC}_{x \rightarrow y} = \frac{\sfrac{\text{RSS}_{y_\text{res}}-\text{RSS}_{y_\text{unr}}}{(p-p^\prime)}}{\sfrac{\text{RSS}_{y_\text{unr}}}{(T-p)}}, \\
\text{GC}_{y \rightarrow x} = \frac{\sfrac{\text{RSS}_{x_\text{res}}-\text{RSS}_{x_\text{unr}}}{(p-p^\prime)}}{\sfrac{\text{RSS}_{x_\text{unr}}}{(T-p)}},
\end{align}
where $\text{RSS}$ means \textit{residual sum of squares}, $p$ and $p^\prime$ are the number of parameters in unrestricted and restricted models, respectively. 
This metric, used to determine Granger causality, quantifies the impact of cross-coupling terms in prediction of a variable. 

The bivariate VAR model in Equation (\ref{eq3}) can also be written as follows: 
\begin{align*}
\vxi_t &= \mA_1\vxi_{t-1}+\mA_2\vxi_{t-2}+\cdots+\mA_{m}\vxi_{t-m} + \vepsilon_t,
\end{align*}
where $\vxi_t := \begin{bmatrix}y(t)\\x(t)\end{bmatrix}$, $\mA_i := \begin{bmatrix}
a_{yy}^{(i)} & a_{yx}^{(i)} \\ a_{xy}^{(i)} & a_{xx}^{(i)}\end{bmatrix}$ and $\vepsilon_t :=\begin{bmatrix}\epsilon_y(t)\\\epsilon_x(t)\end{bmatrix}$. The $m$-th order VAR process can be written as a first order VAR process VAR(1) by stacking the variables as:
\small{
\begin{align*}
\begin{bmatrix}
\vxi_t \\ \vxi_{t-1} \\ \vxi_{t-2} \\ \vdots \\ \vxi_{t-m+1}\end{bmatrix} &= \begin{bmatrix}
\mA_1 & \mA_2 & \cdots &  \mA_{m-1} & \mA_{m} \\
\mId & \mzero & \cdots & \mzero & \mzero \\
\mzero & \mId  & \cdots & \mzero & \mzero \\
\vdots & \vdots & \vdots & \vdots & \vdots \\
\mzero & \mzero & \cdots & \mId  & \mzero \\
\end{bmatrix} \begin{bmatrix}
\vxi_{t-1} \\ \vxi_{t-2} \\ \vxi_{t-3} \\ \vdots \\ \vxi_{t-m}\end{bmatrix} + \begin{bmatrix}
\vepsilon_t \\ \mzero \\ \mzero \\ \vdots \\ \mzero\end{bmatrix}, \\
\vzeta_t &= \mGamma(\mA)\vzeta_{t-1} + \vnu_t,
\end{align*}}
\normalsize 

\noindent 
where $\mGamma(\cdot) : \R^{2 \times 2m} \rightarrow \R^{2m \times 2m}$ is the transformation to extract the companion matrix corresponding to $\mA$. For the VAR model to be stationary, the eigenvalues of $\mGamma(\mA)$ must lie inside the unit circle \cite{shojaie2022granger}.

\section{Proposed Method}



In this paper, we introduce a novel way to augment stationarity constraint with model identification by imposing sparsity on autoregressive coefficients in a hierarchical manner. We assume prior knowledge of the maximum allowable model order denoted as $\bar{m}$ i.e. $m < \bar{m}$. The hierarchical sparsity pattern is created by incorporating the Latent Overlapping Group (LOG) lasso penalty, as detailed in \cite{liu2020fitting,jacob2009group}, into the optimization objective function. Let $\vc = [a_{yy}^{(1)} \cdots a_{yy}^{(\bar{m})}a_{yx}^{(1)} \cdots a_{yx}^{(\bar{m})}a_{xy}^{(1)} \cdots a_{xy}^{(\bar{m})}a_{xx}^{(1)} \cdots a_{xx}^{(\bar{m})}]^\Trsp \in \R^{4\bar{m} \times 1}$ be the vector containing all of the parameters of the VAR model. Then the LOG penalty function is defined as:
\begin{align*}
\Omega_{\text{LOG}}(\vc) = \min_{\vl^{(g)}, g \in \loG}\{\sum_{g \in \loG}w_g\Vert \vl^{(g)}\Vert_2 \mid \sum_{g \in \loG} \vl^{(g)} = \vc,  \vl_{g^c}^{(g)} = 0\},
\end{align*}
where $\loG = \{\{1\},\cdots\{1,..,\bar{m}\},\{\bar{m}+1\},\cdots\{\bar{m}+1,..,2\bar{m}\},\\\{2\bar{m}+1\},\cdots\{2\bar{m}+1,..,3\bar{m}\},\{3\bar{m}+1\},\cdots\{3\bar{m}+1,..,4\bar{m}\} \}$
is the set  of all groups, $\vl^{(g)}  \in \mathbb{R}^{4\bar{m} \times 1}$ is a latent vector indexed by $g$, and $w_g$ is the weight for set $g$. The key concept is that the sparsity-inducing penalty maintains the hierarchical structure of non-zero parameters. For instance, it avoids setting the first AR parameter to zero while the second AR parameter remains non-zero. As a result of $\Omega_{\text{LOG}}(\vc)$, model order for $a_{yy},a_{yx},a_{xy},a_{xx}$ will be chosen adaptively rather than using a fixed order. 

We propose to couple this hierarchical-sparsity based penalty with the stationarity requirement to ensure stability of the model. Our proposed approach incorporates a spectral norm-based regularization technique aimed at constraining the eigenvalues of the companion matrix linked to the estimated parameters to reside within the unit circle. The spectral norm of a matrix is defined as its largest singular value:
\begin{align*}
\Vert \mGamma(\mA) \Vert_2 &:= \max_i \mid \vsigma_i \mid.
\end{align*}
Hence, the condition for stationarity is $\Vert \mGamma(\mA) \Vert_2 < 1$. The proximal  operator $\Psi_{\text{SP}}(\mA)$ for spectral norm can be written as follows:
\begin{align*}
\Psi_{\text{SP}}(\mA) &:= \arg \min_{\mA} \gamma \Vert \mGamma(\mA) \Vert_2 + \frac{\rho}{2} \Vert \mX - \mGamma(\mA) \Vert_F^2.
\end{align*}
The above problem amounts to finding the proximal operator for $\ell_{\infty}$ norm. In particular,   if $\mX= \mU\text{diag}(\vsigma)\mV^\Trsp$ then
$\Psi_{\text{SP}}(\mA)$ can estimated as
$$\Psi_{\text{SP}}(\mA) \approx   \mGamma^{-1}(\mU\text{diag}(\vv^\ast)\mV^\Trsp),$$
where $\mGamma^{-1}(\cdot) : \R^{2\bar{m} \times 2\bar{m}} \rightarrow \R^{2 \times 2\bar{m}} $ is the inverse of $\mGamma$,
\begin{align*}
\vv^\ast  &= \arg \min_{\vv} \sfrac{\gamma}{\rho} \Vert \vv \Vert_\infty + \frac{1}{2} \Vert \vv - \vsigma\Vert_2^2, \\
\Pi_{\sfrac{\gamma}{\rho}}(\mX) &:=  \mGamma^{-1}(\mU\text{diag}(\vv^\ast)\mV^\Trsp),
\end{align*}
where the above operations are combined within $\Pi(\cdot)$ for brevity, and $\loP_{\Vert \cdot \Vert_1 \leq 1}$ is the operator for projection inside the unit $\ell_1$ norm ball, 
which can be computed in a non-iterative fashion by using the approach in \cite{wang2013projection}. 

In addition to reconstruction error, the optimization program will include the $\Omega_{\text{LOG}}$ and $\Psi_{\text{SP}}$ terms to enforce sparsity and stationarity respectively:
\begin{align}
&\notag\min \frac{1}{2} \Vert \mY-\mA\mH\Vert_F^2 + \lambda \Omega_{\text{LOG}}(\vc) +\gamma\Vert \mGamma(\mZ) \Vert_2\\\notag&+\frac{1}{2} \Vert \mY^\prime-\mA^\prime\mH^\prime\Vert_F^2+ \lambda^\prime \Omega_{\text{LOG}}(\vc^\prime)+\gamma^\prime\Vert \mGamma(\mZ^\prime) \Vert_2\\
&\label{op1}\text{s.t.} \hspace*{0.3cm} \vc=\text{vec}(\mA^\Trsp), \hspace*{0.3cm} \vc^\prime=\text{vec}(\mA^{\prime^\Trsp}), \hspace*{0.3cm} \mA=\mZ,\hspace*{0.3cm} \mA^\prime=\mZ^\prime. \\
&\notag\min \frac{1}{2} \Vert \mY-\mA\mH\Vert_F^2 + \lambda \Omega_{\text{LOG}}(\vc) +\gamma\Psi_{\text{SP}}(\mZ)\\
&\label{op2}\text{s.t.} \hspace*{0.3cm} \vc=\text{vec}(\mA^\Trsp), \hspace*{0.3cm} \mA=\mZ.
\end{align}

In contrast to the method proposed in \cite{liu2020fitting}, which employs an iterative minimization approach to achieve a Euclidean projection within a feasible set, we introduce specific constraints on the companion matrix of the model parameters to promote stable estimates by penalizing unstable estimates in the optimization program (\ref{op1}). By expanding $\Omega_{\text{LOG}}(\vc)$ in (\ref{op1}) we get:
\small
\begin{align*}
&\min \frac{1}{2} \Vert \mY-\mA\mH\Vert_F^2+\frac{1}{2} \Vert \mY^\prime-\mA^\prime\mH^\prime\Vert_F^2 +\gamma\Vert \mGamma(\mZ) \Vert_2\\&+\gamma^\prime\Vert \mGamma(\mZ^\prime) \Vert_2+ \lambda \sum_{g \in \loG}w_g \Vert \mP_{.g}\Vert_2+\frac{1}{2} \Vert \sum_{g \in \loG} \mQ_{.g}-\vc \Vert_2^2 \\&+ \lambda^\prime \sum_{g \in \loG}w_g \Vert \mP_{.g}^\prime\Vert_2+\frac{1}{2} \Vert \sum_{g \in \loG} \mQ_{.g}^\prime-\vc^\prime \Vert_2^2\\
&\text{s.t.} \\&\hspace*{0.008cm} \vc=\text{vec}(\mA^\Trsp),\hspace*{0.008cm} \mP=\mQ, \hspace*{0.008cm} \vc^\prime=\text{vec}(\mA^{\prime^\Trsp}),\hspace*{0.008cm} \mP^\prime=\mQ^\prime, \hspace*{0.008cm} \mA=\mZ,\\& \mA^\prime=\mZ^\prime, \hspace*{0.008cm} (\mP_{.g})_{g^c} = 0, \hspace*{0.008cm} a_{yx}^{(j)^\prime} = a_{xy}^{(j)^\prime} = 0 \hspace*{0.3cm}\forall j=1,\cdots,\bar{m}.
\end{align*}
\normalsize The update for parameter matrix $\mA$ can be computed by finding the gradient of objective with respect to $\mA$ as follows:
\begin{align*}
\mA&= [\rho_3\tilde{\mZ}+\rho_1{\text{vec}^{-1}(\tilde{\vc})}^\Trsp)+\mY\mH^\Trsp](\mH\mH^\Trsp + (\rho_1+\rho_3)\mId)^{-1},
\end{align*}
where $\tilde{\mZ}:=\mZ-\mU_3$, $\tilde{\vc}:=\vc-\vu_1$ $\rho := \rho_1+\rho_3$. To estimate $\mA^\prime$ for restricted model, there is an additional constraint $a_{xy}^{(i)^\prime} = a_{yx}^{(i)^\prime} = 0 \hspace*{0.3cm}\forall i=1,\cdots,\bar{m}$. To address this assumption of known support, two approaches can be employed: i) first, by increasing the hierarchical sparsity parameter $\lambda^\prime$, especially on cross-coupling terms, to force them to become zero, ii) second, by discarding the cross-coupling terms in each iteration after solving the ordinary least squares problem. To avoid difficulties associated with both approaches, we use the idea to shift the sparsity from $\mA^\prime$ to data matrices $\mY$ and $\mH$ such that cross-coupling terms do not play any role in optimization for \textit{restricted} model. By redefining matrices in Equation (\ref{eq3}) we can write:
\begin{align*}
\mY^\prime = \mA^\prime\mH^\prime + \mE^\prime,
\end{align*}
where $\mY^\prime := \begin{bmatrix} y(t) & \vzero \\ \vzero & x(t)\end{bmatrix} \in \R^{2 \times 2T^\prime}$, $\mH^\prime := \left[\begin{smallmatrix}y(t-1) & \\ \vdots & \mzero \\ y(t-\bar{m}) &  \\  & x(t-1) \\ \mzero & \vdots \\ & x(t-\bar{m})\end{smallmatrix}\right] \in \R^{2\bar{m} \times 2T^\prime}$,  $\mE^\prime := \begin{bmatrix} n_y(t)^\prime & \vzero \\ \vzero & n_x(t)^\prime\end{bmatrix} \in \R^{2 \times 2T^\prime}$, and $\mA^\prime := \begin{bmatrix}a_{yy}^{(1)^\prime} \cdots a_{yy}^{(\bar{m})^\prime} & a_{yx}^{(1)^\prime}  \cdots a_{yx}^{(\bar{m})^\prime} \\ a_{xx}^{(1)^\prime} \cdots a_{xx}^{(\bar{m})^\prime} & a_{xy}^{(1)^\prime}  \cdots a_{xy}^{(\bar{m})^\prime}\end{bmatrix} \in \R^{2 \times 2\bar{m}}$. Now we can update $\mA^\prime$ according to: 
\begin{align*}
\mA^\prime&= [\rho_3^\prime\tilde{\mZ^\prime}+\rho_1{\text{vec}^{-1}(\tilde{\vc^\prime})}^\Trsp)+\mY^\prime\mH^{\prime\Trsp}](\mH^\prime\mH^{\prime\Trsp} + \rho^\prime\mId)^{-1},
\end{align*}
where $\rho^\prime := \rho_1^\prime+\rho_3^\prime$. After computing $\mA^\prime$ we can safely discard cross-coupling terms without any loss of information. Closed-form solutions for hierarchical sparsity terms $\mP, \mQ$ and $\mP^\prime,\mQ^\prime$ are obtained as discussed in \cite{liu2020fitting}. For details, refer to Algorithm 1.
To find the update for $\mZ$, 
we leverage the following two properties of $\mGamma(\cdot)$ operator.
\subsubsection*{Property 1}\label{P1} $\Vert \mGamma(\mX) \Vert_F^2 = \Vert \mX \Vert_F^2 + 2(\bar{m}-1)$ for any $\mX \in \R^{2 \times 2\bar{m}}$.
\subsubsection*{Property 2}\label{P2}  $\mGamma(\mX - \mY) = 2\mGamma(\frac{1}{2}\mX)-\mGamma(\mY)$ for any $\mX$, $\mY \in \R^{2 \times 2\bar{m}}$.

\noindent These two properties allow us to find
 the closed-form solution for the update for $\mZ$ as: 
\begin{align*}
\mZ &= \arg \min_{\mZ} \gamma\Vert \mGamma(\mZ)  \Vert_2 + \frac{\rho_3}{2} \Vert \mA+\mU_3-\mZ\Vert_F^2, \\
&=\arg \min_{\mZ} \gamma\Vert \mGamma(\mZ)  \Vert_2 + \frac{\rho_3}{2} \Vert 2 \mGamma( \frac{1}{2}(\mA+\mU_3))-\mGamma(\mZ)\Vert_F^2, \\
\mZ&= \Pi_{\sfrac{\gamma}{\rho_3}} (2 \mGamma( \frac{1}{2}(\mA+\mU_3))).
\end{align*}
Unrestricted and restricted models are evaluated by Algorithm 1. 
\begin{algorithm}[t]
\caption{SS-ADMM for GC Estimation}
\textbf{Input:} $\mY,\mH,\bar{m},\lambda, \gamma, \rho_1, \rho_2, \rho_3$
	\begin{algorithmic}
		\While {stopping criterion not met}
			\State $\mB := \rho(\mZ-\mU_3+{\text{vec}^{-1}(\vc -\vu_1)}^\Trsp)$
			\State $\mA = (\mB+\mY\mH^\Trsp)(\mH\mH^\Trsp + \rho\mId)^{-1}$
			\vspace*{-0.4cm}\State \For{$g = 1,...,4\bar{m}$} 
				\State $\mP_{gg} = \prox_{\lambda w_g \Vert \cdot \Vert }(\mP_{gg} + \vq_g - \vu_{2_g} -\vp_g)$
				\State $\mP_{g^cg} = 0$
				\EndFor
			\State $\vp = \sfrac{1}{4\bar{m}} \sum_{g \in \loG} \mP$
			\State $\vq = \sfrac{1}{(\rho_2+4\bar{m})} (\vc + \rho_2(\vu_2+\vp))$
			\State $\vc = \text{vec}(\mA^\Trsp)+ \sum_{g \in \loG} \mP+\text{vec}(\mU_1)$
			\State	$\mZ = \mPi_{\sfrac{\gamma}{\rho_3}} (2 \mGamma( \frac{1}{2}(\mA+\mU_3))) $		
			\State $\vu_1 = \vu_1 + \text{vec}(\mA^\Trsp)-\vc$
			\State $\vu_2 = \vu_2 + \vp - \vq$
			\State $\mU_3 = \mU_3 + \mA - \mZ$
			\State $m_i = \text{card}(\mP_{(i-1)\bar{m}+1:i\bar{m}}\neq 0)$  $\forall i=1,2,3,4$
		\EndWhile
	\end{algorithmic}
\textbf{Output:} $\mA,m_1,m_2,m_3,m_4$ 
\end{algorithm} 
The calculation of inverse for the estimate of $\mA$ is computationally expensive, therefore, Cholesky decomposition has been used to pre-compute the inverse of the matrix once. For initialization phase one-time cost is $\mathcal{O}(8\bar{m}^3+4T\bar{m})$, and the total computational cost for the iterative process is  $\mathcal{O}(8\bar{m}^2+8\bar{m}^3+2\bar{m}\log(2\bar{m})) \times $ number of iterations.

\section{Wavelet Thresholding for non-Gaussian excitation noise}

In the above-mentioned model the excitation noise term represents the process noise associated with inherent dynamics of the system being modeled. In the context of Granger causality estimation based on an autoregressive 
model, one of the primary assumptions is that the variables need to be observed without measurement errors \cite{shojaie2022granger}. For simplicity, most of the analysis methods assume that the measurement noise is negligible and the recorded data represent the true underlying process dynamics. However, this assumption is technically incorrect and, therefore, leads to significantly biased estimate of the model and false inference of causation when the underlying model is affected by considerable noise \cite{nalatore2007mitigating, abbas2024infr}. Hence, handling measurement noise in the model is crucial to obtain reliable Granger causality estimates. Estimating Granger causality in the presence of measurement errors has scarcely been investigated in the literature. Existing approaches mostly rely on Kalman filtering and Expectation Maximization \cite{nalatore2007mitigating}, \cite{ccayir2021maximum}, \cite{park2019measurement} techniques using a state space representation of the model. However, Kalman filtering assumes the knowledge of an exact mathematical model of the system and Gaussian assumption for both the state and measurement noise. Deviations from these assumptions can lead to suboptimal performance. Expectation Maximization, on the other hand, is sensitive to the choice of initial parameters and involves iterative optimization steps resulting in high computational complexity, especially for large and complex systems. 

To overcome the limitations associated with these approaches, leveraging empirical knowledge about the recorded signals can result in a realistic model for incorporating measurement noise. In well-controlled experimental settings with high-quality EEG equipment, where the noise characteristics are relatively stable over time, it is reasonable to assume the measurement noise to be stationary Gaussian noise. The physiological noise, on the other hand, can be represented in the form of non-Gaussian trends such as skewness or sudden bursts of very large amplitudes at random times.

In the context of EEG and sEMG signals, observing such non-Gaussian trends and measurement noise is unavoidable.
Therefore, to account for this, instead of assuming Gaussianity on the excitation noise sequence as a whole, we propose to assume Gaussianity on the recorded measurement noise and the excitation noise term containing non-Gaussian trends is assumed to be sparse in a wavelet basis. This assumption offers a practical approach to modeling real signals, as supported by previous research in \cite{averkamp2003wavelet} and \cite{antoniadis2002wavelet}. As a result, it is expected to obtain improved causality estimates.

Revisiting the bivariate VAR model in Equation (\ref{eq3}), now we consider the presence of measurement noise in the observed signals. The observed signals $\hat{\mY}$ and their past values $\hat{\mH}$ can, therefore, be represented as a sum of measurement noise-free signal and an additive Gaussian measurement noise term, $\hat{\mY}={\mY}+\mDelta \mY$ and $\hat{\mH}={\mH}+\mDelta \mH$, respectively.
From the original autoregressive model in (\ref{eq3}) it follows then that the observed signals are related according to:
 \small{
\begin{align}
\hat{\mY} - \mDelta\mY &= \mA (\hat{\mH}-\mDelta\mH) + \mE, \label{eq9} \\
\hat{\mY}  &= \mA \mH + \mE +\mDelta\mY. \label{eq8}
\end{align}}\normalsize
The model in Equation (\ref{eq9}) is ill-posed in a conventional setting, since both the excitation noise $\mE$ and the measurement noise $\mDelta\mY$ and $\mDelta\mH$ are, traditionally, assumed to be Gaussian. However, considering a realistic signal model we propose to enforce a wavelet sparsity assumption on the excitation noise $\mE$, whereas the measurement noise terms $\mDelta\mY$ and $\mDelta\mH$ are assumed to be Gaussian and relatively weak compared to the physiological signal. By using these constraints, we formulate a non-convex optimization problem based on the model proposed in Equation (\ref{eq8}). 

We formulate the objective in terms of $\mH$ that will eventually be computed by subtracting the noise values. 
\begin{figure}[t]
    \centering
    \includegraphics[width=0.6\linewidth]{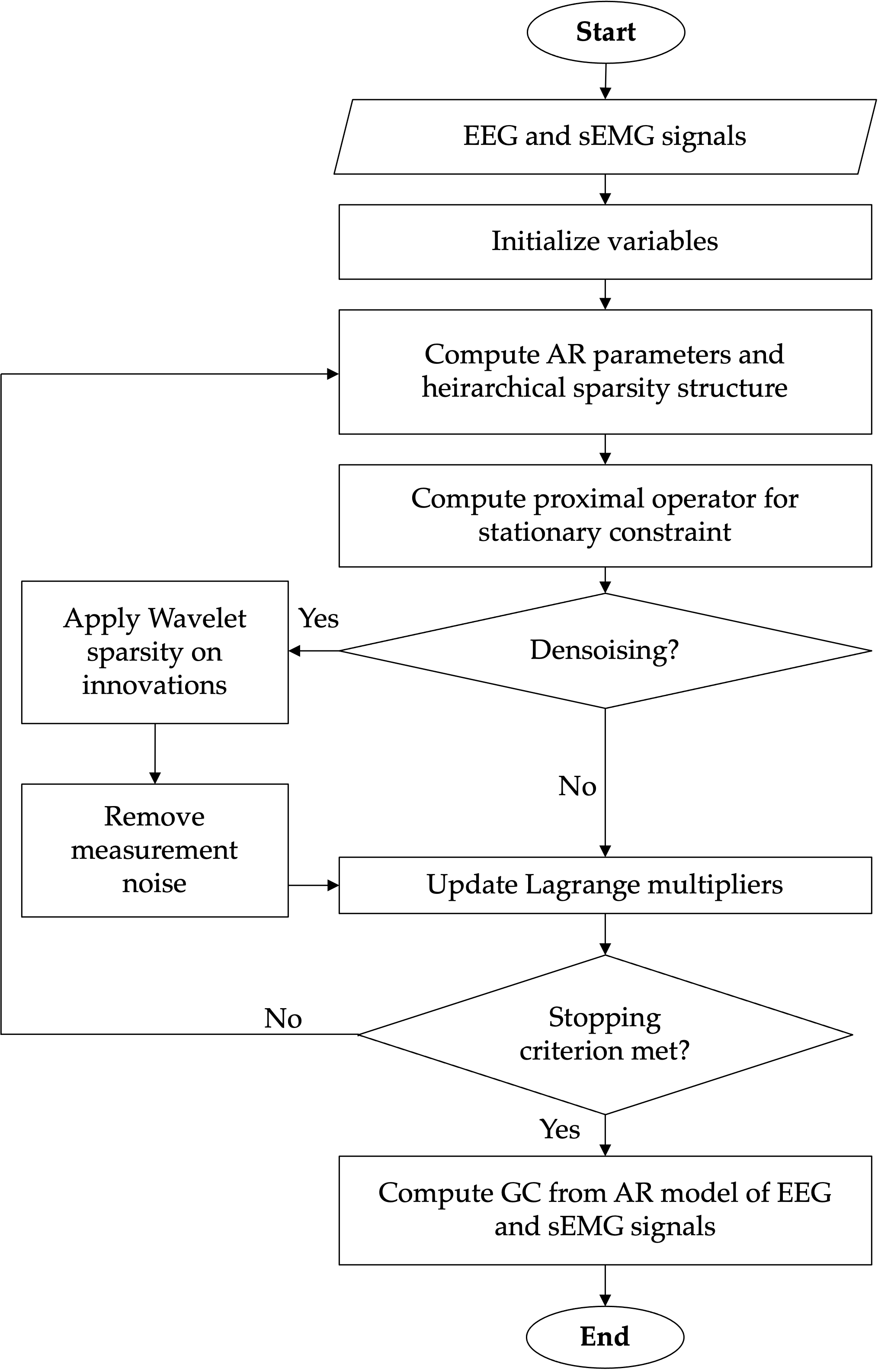}
    \caption{{Proposed method overview: with (SSD-ADMM) vs. without (SS-ADMM) denoising.}}
    \label{fig:enter-label}
\end{figure}
The overall objective function can be written as follows:
\small{
\begin{align*}
&\min \kappa \Vert \setW(\mE)\Vert_1+\frac{\alpha}{2}\Vert \hat{\mY}  -\mA \mH - \mE\Vert_F^2+\lambda \Omega_{\text{LOG}}(\vc) +\gamma\Psi_{\text{SP}}(\mZ)\\
&\text{s.t.}  \hspace*{0.3cm} \vc=\text{vec}(\mA^\Trsp), \hspace*{0.3cm} \mA=\mZ ,
\end{align*}}\normalsize
where $\set{W}(\cdot)$ indicates an orthogonal wavelet transform. In the above constrained optimization problem, the last two terms enforce hierarchical sparsity and stationarity, as in (\ref{op2}). 

The above program is non-convex due to the presence of bilinear term $\mA\mH$. However, the matrix $\mH$ has a special Toeplitz structure that can be leveraged to efficiently handle the non-convex bilinear term. The augmented Lagrangian of the above program in scaled form, can be written as follows:
\small{
\begin{align*}
\min& \kappa \Vert \setW(\mE)\Vert_1+\frac{\alpha}{2}\Vert \hat{\mY}  -\mA \mH - \mE\Vert_F^2+\lambda \Omega_{\text{LOG}}(\vc) +\gamma\Psi_{\text{SP}}(\mZ)\\&+\frac{\rho_1}{2}\Vert \mA-\text{vec}^{-1}(\vc-\vu_1)\Vert_F^2+\frac{\rho_3}{2}\Vert \mA-\mZ+\mU_3\Vert_F^2,
\end{align*}}\normalsize
which can be solved using non-convex ADMM.
The closed form solutions for all the matrices can be obtained by fixing the other variables and optimizing with respect to one of them. The closed form solutions to enforce hierarchical sparsity and stationarity conditions are estimated by solving (\ref{op1}).
By keeping all other matrices fixed, the update for parameter matrix $\mA$ can be computed as follows:
\small{\begin{align}
\notag
\mA &= \arg \min_{\mA}\frac{\alpha}{2}\Vert \hat{\mY}  -\mA \mH - \mE\Vert_F^2 +  \frac{\rho_3}{2} \Vert \mA-\mZ+\mU_3 \Vert_F^2 \\&+\frac{\rho_1}{2} \Vert \mA-\text{vec}^{-1}(\vc-\vu_1)^T\Vert_F^2, \\
\notag
&= [\rho_3 (\tilde{\mZ})+\rho_1(\text{vec}^{-1}(\tilde{\vc})^T)+\alpha\tilde{\mY} \mH ^T][ \alpha \mH\mH ^T+(\rho_3+\rho_1)\mI]^{-1},
\end{align}}\normalsize
where $\tilde{\mZ}:=\mZ-\mU_3$, $\tilde{\vc}:=\vc-\vu_1$, and $\tilde{\mY}:=\hat{\mY} -\mE$. The update for excitation noise matrix $\mE$ can be simplified as follows:
\small{\begin{align*}
\mE &= \arg \min_{\mE}\kappa \Vert \setW(\mE)\Vert_1+\frac{\alpha}{2}\Vert \hat{\mY}  -\mA\mH - \mE\Vert_F^2, \\
&= \setW^{-1}(\setS_{\sfrac{\kappa}{\alpha}}(\setW(\hat{\mY}-\mA\mH)),
\end{align*}}\normalsize
where $\setW^{-1}(\cdot)$ is the inverse of the orthogonal wavelet transform and $\setS(\cdot)$ is the soft-thresholding operator. The update for the measurement noise matrix $\mDelta\mY$  turns out to be:
\small{\begin{align*}
\mDelta\mY = \hat{\mY}-\mA\mH -\mE~.
\end{align*}}\normalsize 
In order to find the update for $\mH$, we first estimate the noise matrix $\mDelta\mH$. After updating all the matrices, $\mDelta\mH$ can be updated as a function of these matrices by leveraging the inherent Toeplitz structure as follows:
\small{\begin{align}
\mDelta\mH = \begin{bmatrix}
n_y(\bar{m})&\cdots&n_y(2\bar{m}-1)&\cdots& n_y(\bar{m}+T^\prime-1) 
\\
n_y(\bar{m}-1)&\cdots&n_y(2\bar{m}-2)&\cdots& n_y(\bar{m}+T^\prime-2)
\\ 
&&\vdots& 
\\
n_y(1)&\cdots&n_y(\bar{m})&\cdots& n_y(T^\prime)
\\
\\
n_x(\bar{m})&\cdots&n_x(2\bar{m}-1)&\cdots& n_x(\bar{m}+T^\prime-1) 
\\
n_x(\bar{m}-1)&\cdots&n_x(2\bar{m}-2)&\cdots& n_x(\bar{m}+T^\prime-2)
\\ 
&&\vdots& 
\\
n_x(1)&\cdots&n_x(\bar{m})&\cdots& n_x(T^\prime)\\
\end{bmatrix},\label{eq11}
\end{align}}\normalsize 
where $\mDelta\mH = \begin{bmatrix} \mT_y \\ \mT_x\end{bmatrix}$, and $\mT_y \in \mathbb{R}^ {\bar{m}\times T^\prime }$ and $\mT_x \in \mathbb{R}^ {\bar{m}\times T^\prime }$ are rectangular Toeplitz matrices containing noise terms in $y(t)$ and $x(t)$, respectively. Due to the Toeplitz structure both $\mT_y$ and $\mT_x$ can be fully determined by their first row and first column. To determine the first row of $\mT_y$ and $\mT_x$, by comparing Equation(\ref{eq9}) and Equation(\ref{eq11}), we observe
\begin{align}
\mT_{y_{1,2:}} &= \mDelta\mY_{1,:} \label{eq13}~, \\
\mT_{x_{1,2:}} &= \mDelta\mY_{2,:} \label{eq14}~.
\end{align}
The first column of $\mDelta\mH = \begin{bmatrix}
\mT_y \\ \mT_x
\end{bmatrix}$ can be determined efficiently by solving a simple least square problem. Consider the Equation (\ref{eq9}) for $t=\bar{m}+1$:
\footnotesize{\begin{align*}
&\begin{bmatrix}
\hat{y}(\bar{m}+1) \\ \hat{x}(\bar{m}+1) 
\end{bmatrix} - \begin{bmatrix}
n_y(\bar{m}+1) \\ n_x(\bar{m}+1)
\end{bmatrix} \\&=
\begin{bmatrix}
a_{yy}^{(1)}\cdots a_{yy}^{(\bar{m})}&a_{yx}^{(1)}\cdots a_{yx}^{(\bar{m})}\\a_{xy}^{(1)}\cdots a_{xy}^{(\bar{m})}&a_{xx}^{(1)}\cdots a_{xx}^{(\bar{m})}
\end{bmatrix}\left(\begin{bmatrix}\hat{y}(\bar{m})\\\hat{y}(\bar{m}-1)\\\vdots \\\hat{y}(1)\\\hat{x}(\bar{m})\\\hat{x}(\bar{m}-1)\\\vdots \\\hat{x}(1)
\end{bmatrix} - \begin{bmatrix}n_y(\bar{m})\\n_y(\bar{m}-1)\\\vdots \\n_y(1)\\n_x(\bar{m})\\n_x(\bar{m}-1)\\\vdots \\n_x(1)
\end{bmatrix} \right)\\&+\begin{bmatrix}\epsilon_y(\bar{m}+1) \\ \epsilon_x(\bar{m}+1)\end{bmatrix} ~~.
\end{align*}}\normalsize 
Assigning the matrices below to their corresponding values in the equation above gives:
\begin{align*}
\hat{\vy} - \mDelta\vy &= \mA (\vh-\mDelta\vh ) + \ve, \\
\mA\mDelta\vh   &= \mA \vh-\hat{\vy} + \ve +\mDelta\vy, \\
\mDelta\vh  &= \mA^T(\mA\mA^T)^{-1}(\mA \vh-\hat{\vy} + \ve +\mDelta\vy )~~.
\end{align*}
Hence we can write
\begin{align}
\mDelta\mH_{:,1} = \begin{bmatrix}
\mT_{y_{:,1}} \\ \mT_{x_{:,1}}
\end{bmatrix} = \mDelta\vh \label{eq12}~~.
\end{align}
The measurement noise matrix $\mDelta\mH$ can then be obtained from equations (\ref{eq13}-\ref{eq14}) and equation(\ref{eq12}) and by exploiting its Toeplitz structure. 
The uncontaminated matrix $\mH$ can then be computed by subtracting estimated noise from the noisy observations as:
\begin{align*}
\mH&= \hat{\mH}-\mDelta\mH.
\end{align*} 
Finally, the Lagrange multipliers are computed and all the variables are updated iteratively.
The extension of SS-ADMM to incorporate measurement noise in recorded signals is termed as SSD-ADMM: Stationary and Sparse Denoising ADMM. The complete method is shown in Algorithm 2.
\begin{algorithm}[t]
\caption{SSD-ADMM for GC Estimation}
\textbf{Input:} $\hat{\mY},\hat{\mH},\bar{m},\lambda, \gamma, \alpha, \kappa, \rho_1, \rho_2, \rho_3$
	\begin{algorithmic}
		\While {stopping criterion not met}
			\State $\mA = [\rho_3 (\mZ-\mU_3)+\rho_1(\text{vec}^{-1}(\vc-\vu_1)^T)+\alpha(\hat{\mY} -\mE)( \mH )^T][ \alpha \mH\mH ^T+(\rho_1+\rho_3)\mI]^{-1}$
			\vspace*{-0.4cm}\State \For{$g = 1,...,4\bar{m}$} 
				\State $\mP_{gg} = \prox_{\lambda w_g \Vert \cdot \Vert }(\mP_{gg} + \vq_g - \vu_{2_g} -\vp_g)$
				\State $\mP_{g^cg} = 0$
				\EndFor
			\State $\vp = \sfrac{1}{4\bar{m}} \sum_{g \in \loG} \mP$
			\State $\vq = \sfrac{1}{(\rho_2+4\bar{m})} (\vc + \rho_2(\vu_2+\vp))$
			\State $\vc = \text{vec}(\mA^\Trsp)+ \sum_{g \in \loG} \mP+\text{vec}(\mU_1)$
			\State	$\mZ = \mPi_{\sfrac{\gamma}{\rho_3}} (2 \mGamma( \frac{1}{2}(\mA+\mU_3))) $	
			\State $\mE = \setW^{-1}(\setS_{\sfrac{\kappa}{\alpha}}(\setW(\hat{\mY}-\mA\mH))$	
			\State $\mDelta\mY = \hat{\mY}-\mA\mH -\mE$
			\State $\mDelta\mH_{\{1,\bar{m}+1\},:} =  \begin{bmatrix} \mDelta\mY_{1,:}  \\
\mDelta\mY_{2,:} \end{bmatrix} $
\State $\mDelta\mH_{:,1} =  \mA^T(\mA\mA^T)^{-1}(\mA \vh-\hat{\vy} + \ve +\mDelta\vy )$
\State $\mH = \hat{\mH}-\mDelta\mH$
			\State $\vu_1 = \vu_1 + \text{vec}(\mA^\Trsp)-\vc$
			\State $\vu_2 = \vu_2 + \vp - \vq$
			\State $\mU_3 = \mU_3 + \mA - \mZ$
			\State $m_i = \text{card}(\mP_{(i-1)\bar{m}+1:i\bar{m}}\neq 0)$  $\forall i=1,2,3,4$
		\EndWhile
	\end{algorithmic}
\textbf{Output:} $\mA, \mY, \mH, \mE, m_1,m_2,m_3,m_4$ 
\end{algorithm}


\section{EXPERIMENTAL RESULTS}
\label{sec:majhead}
In this section, we perform a comparative study to investigate the performance of proposed methods Stationary and Sparse ADMM (SS-ADMM) and its extension Stationary and Sparse Denoising ADMM (SSD-ADMM) against existing methods on both synthetic and real datasets. All the experiments have been performed in MATLAB 2022b with Core i7 CPU (2.90 GHz), 8 GB RAM, and Windows 11 operating system.
\subsection{Results on Synthetic Data}
\begin{figure}[h!tbp]
\begin{minipage}[b]{0.9\linewidth}
  \centering
  \centerline{\includegraphics[width=1.0\textwidth]{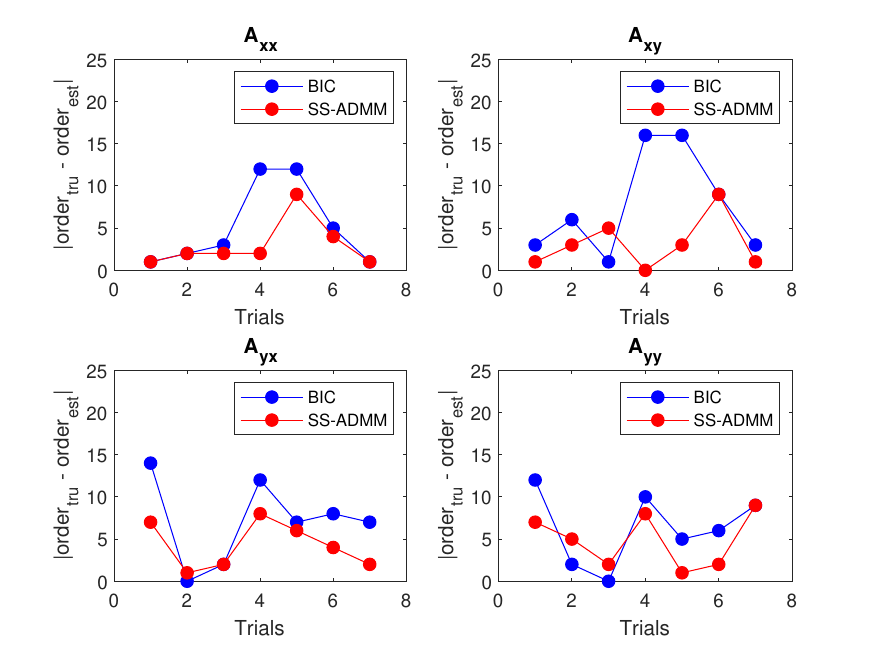}}
\end{minipage}
\caption{\textit{Model order identified by SSD-ADMM Vs BIC \cite{schwarz1978estimating}.}}
\label{Fig1}
\end{figure}

Synthetic data has been generated using an AutoRegressive Fractionally Integrated Moving Average (ARFIMA) model with long memory as outlined in \cite{box2015time}. The bivariate process has $942$ samples where the model orders for $a_{yy},a_{yx},a_{xy}$ and, $a_{xx}$ are set to be $17,21,20$ and, $18$ respectively. We assume that the upper limit on the model orders $\bar{m}$ is $30$. The measurement noise variance is set to be $0.01$. System parameters have been generated randomly based on an underlying mathematical model that governs variables $\mY$, $\mA$, and $\mH$. The results, obtained through multiple experimental repetitions, are compared to evaluate performance in terms of stationarity, sparsity and denoising. The performance of model identification is depicted in Figure \ref{Fig1}, comparing it with BIC \cite{schwarz1978estimating}, which demonstrates that the proposed method achieves more accurate system order estimation. We observed also (not shown in the figure) that SS-ADMM\footnote[1]{{The implementation code will be publicly available.}} tends to estimate the model order as either equal to or higher than the true order, while BIC typically underestimates the model order, thereby losing significant information about the underlying process. 

It is worth highlighting that SS-ADMM offers flexibility by allowing for fine-tuning of the hyperparameter $\lambda$ to precisely estimate the model order. {We performed grid search method \cite{shekar2019grid} to find the optimal set of hyperparameter values where the domain of the hyperparameters is divided into a discrete grid and for every combination of values in the grid the performance metrics are computed using cross-validation.
The combination of values that maximizes the average value in cross-validation is eventually selected.} The proposed method also distinguishes between four distinct model orders for self-coupling and cross-coupling in a bivariate model, as opposed to just two in the case of BIC. 

\vspace{-0.5cm}
\begin{figure}[h!tbp]
\begin{minipage}[b]{0.9\linewidth}
  \centering
  \centerline{\includegraphics[width=1.05\textwidth]{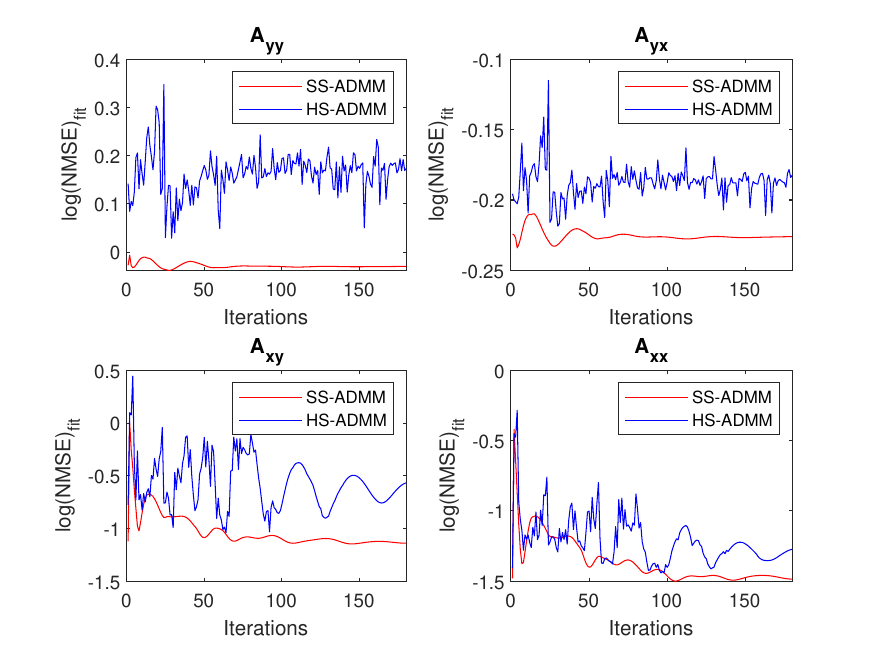}}
\end{minipage}
\caption{\textit{Log of normalized error of HS-ADMM \cite{liu2020fitting} vs SS-ADMM.}}
\label{Fig2}
\end{figure}

Next we evaluate our method in comparison with Hierarchical Sparsity ADMM (HS-ADMM) \cite{liu2020fitting} in terms of computational efficiency.
The results obtained through the enforcement of stationarity using Euclidean projections, as demonstrated in \cite{liu2020fitting}, are contrasted with our proposed regularization based method, $\Psi_{\text{SP}}$, in Figure  \ref{Fig2}. Notably, the projection-based approach exhibits a longer execution time. Conversely, our proposed SS-ADMM approach achieves a speedup of $2\times$ or more per iteration, and the model-fitting error reduces more steadily when executed for the same number of iterations, as illustrated in Figure \ref{Fig2}. The total execution time for HS-ADMM was 1.9 minutes whereas SS-ADMM only took 0.39 seconds.

\begin{figure}[t]
\centering
\subfigure[\textit{\footnotesize{Model Fit}}]
{\includegraphics[trim={1cm 1cm 1cm 2.5cm},clip,height = 3.5cm, width = 4.2cm]{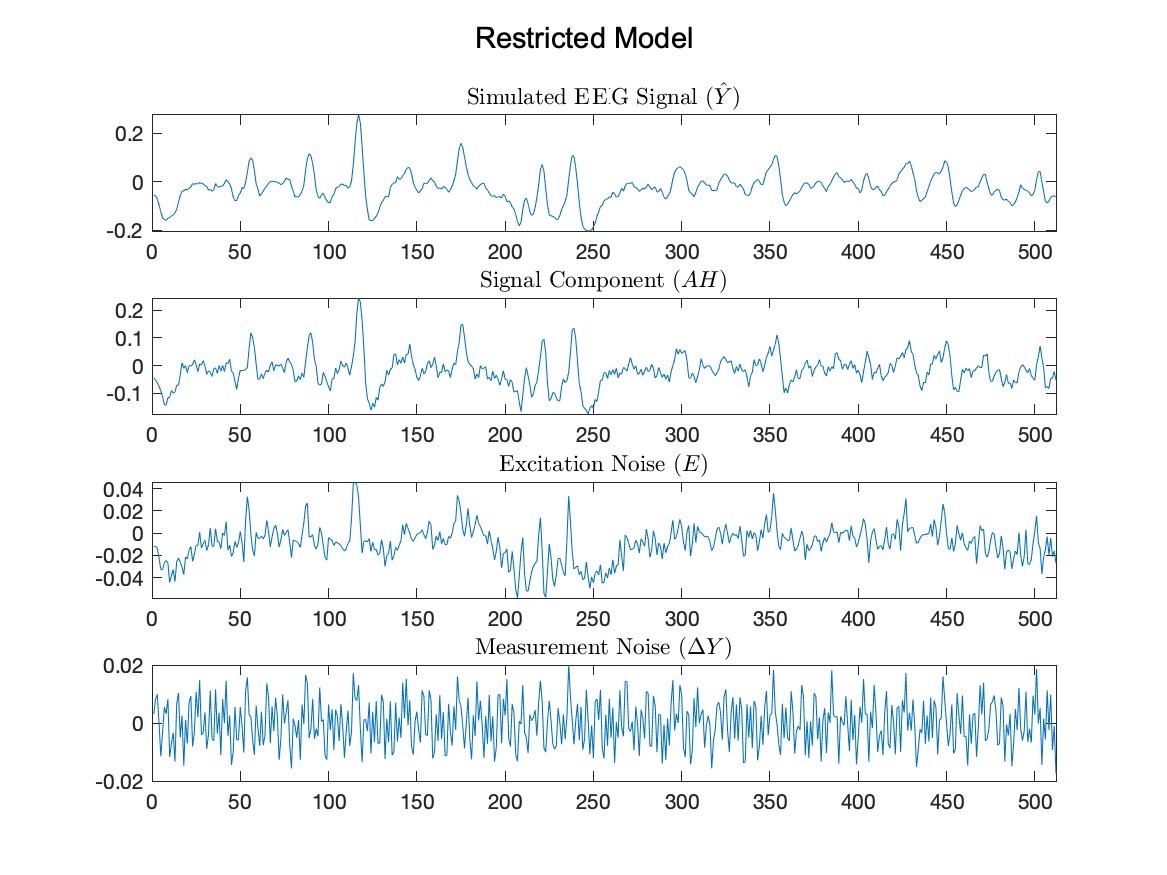}}
\subfigure[\textit{\footnotesize{NMSE Vs $var(\mDelta\mY)$}}]{\includegraphics[trim={0cm 0cm 0cm 0cm},clip,height = 3.7cm, width = 4.5cm]{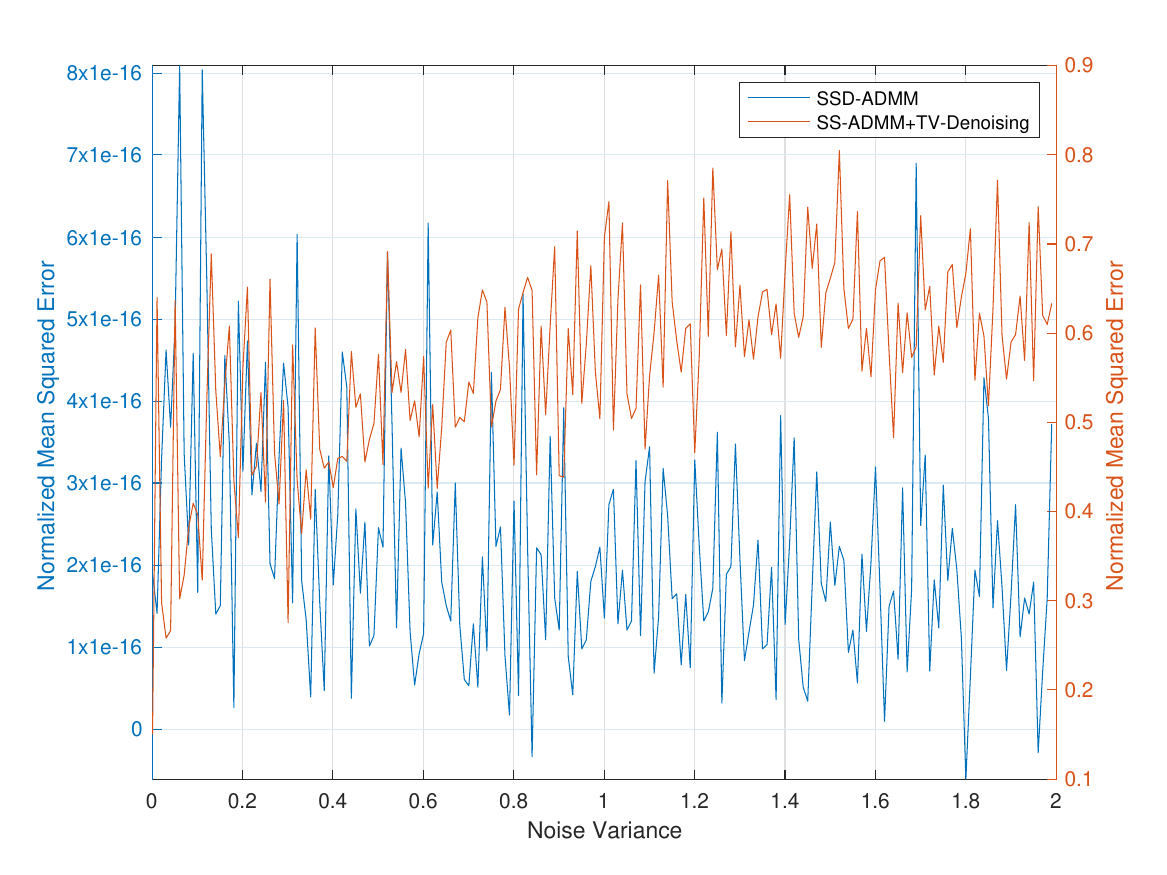}}
\caption{\textit{Model estimation and noise robustness for simulated EEG signal by SSD-ADMM}}
\label{Fig3}
\end{figure}

To observe the denoising performance of of SSD-ADMM, a measurement noise signal sampled from a multivariate normal distribution has been added to the simulated EEG and EMG signals. The noise variance $\sigma_{\mDelta\mY}$ is varied from $0$ to $1.41$. A sensing dictionary has been generated using a discrete Daubechies wavelet transform. A randomly generated sparse vector, representing predominantly zero coefficients in the wavelet basis, is then subject to transformation with a sub-matrix derived from the dictionary, resulting in a wavelet sparse signal for modelling the excitation noise. The sparsity level is set to $70\%$. According to the model discussed in Equation (\ref{eq8}), the observed noisy signal $\hat{\mY}$ can be decomposed as a combination of an autoregressive signal component $\mA\mH$, an excitation noise $\mE$ and a measurement noise term $\mDelta\mY$. The estimates obtained for fitting the proposed model for simulated data as shown in Figure \ref{Fig3}(a).

In Figure \ref{Fig3}(b), it can be observed that if signals are denoised first using a standard method such as Total Variation denoising \cite{vogel1996iterative}, for instance, and then modelled using SS-ADMM, it experiences a degradation in the quality of fit quantified by Normalised Mean Squared Error (NMSE) because of the its inability to handle measurement noise in data. On the other hand, SSD-ADMM effectively accounts for the measurement noise to estimate the clean signals and, thus, yields a more robust model.

\subsection{Results on Physiological Data}
\label{ssec:subhead}
The proposed method is tested on physiological data collected from eight healthy subjects in a previously published study \cite{mcclelland2012modulation}. {Ethical approval was obtained from the Riverside Research Ethics Committee, London, UK.} The subjects performed a controlled motor task, grasping a ruler between thumb and index finger of the right hand. An electromechanical tapper provided mechanical perturbations of lateral displacement to the ruler at pseudorandom intervals of 5.6 - 8.4 s (mean 7 s). The experiment comprised 8 blocks of 25 trials. sEMG was recorded from first dorsal interosseous and bipolar scalp EEG was recorded over left sensorimotor cortex. Both signals were sampled at 1024 Hz, amplified and band-pass filtered (0.5-100 Hz for EEG; 5-500 Hz for sEMG). Offline, data were divided into 5 s epochs (1.1 s pre- and 3.9s post-stimulus). Epochs containing movement or blink artefacts were eliminated. 

\begin{figure}[t]
\centering
\subfigure[\textit{\small Subject B}]{\includegraphics[trim={0.2cm 1cm 1.5cm 1cm},clip,height = 4.6cm, width = 2.1cm]{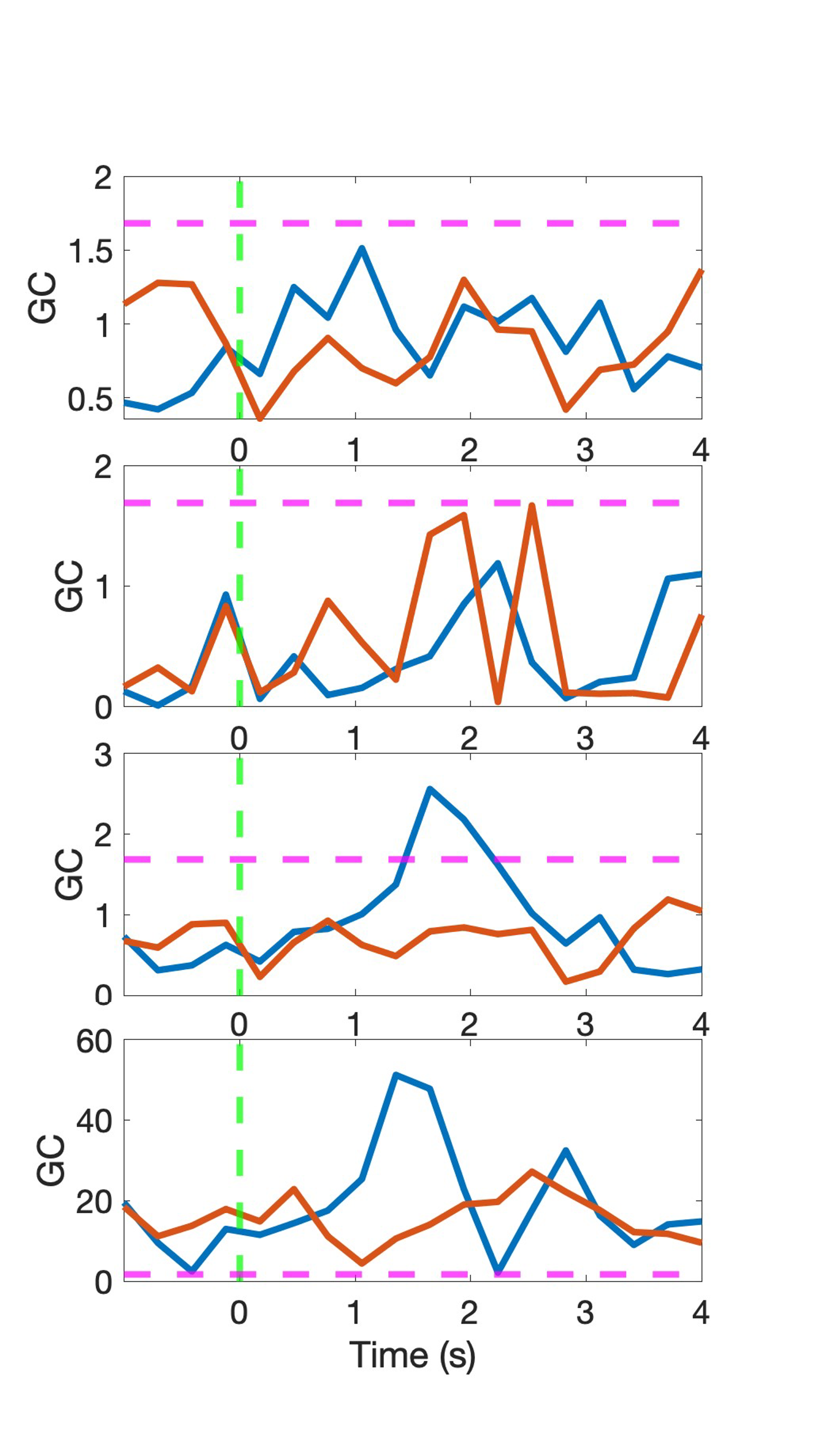}}
\subfigure[\textit{\small Subject D}]{\includegraphics[trim={0.2cm 1cm 1.5cm 1cm},clip,height = 4.6cm, width = 2.1cm]{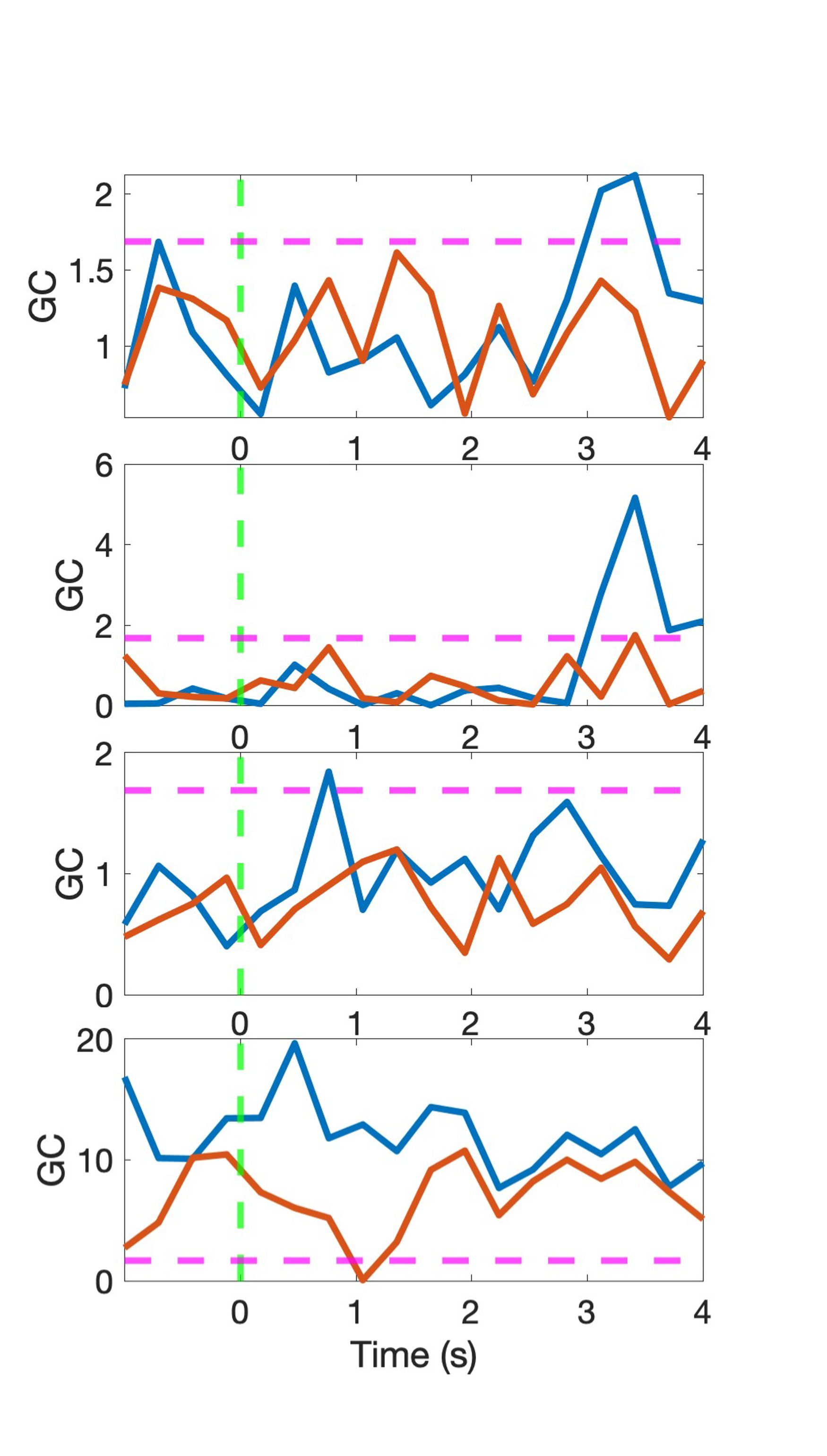}}
\subfigure[\textit{\small Subject G}]{\includegraphics[trim={0.2cm 1cm 1.5cm 1cm},clip,height = 4.6cm, width = 2.1cm]{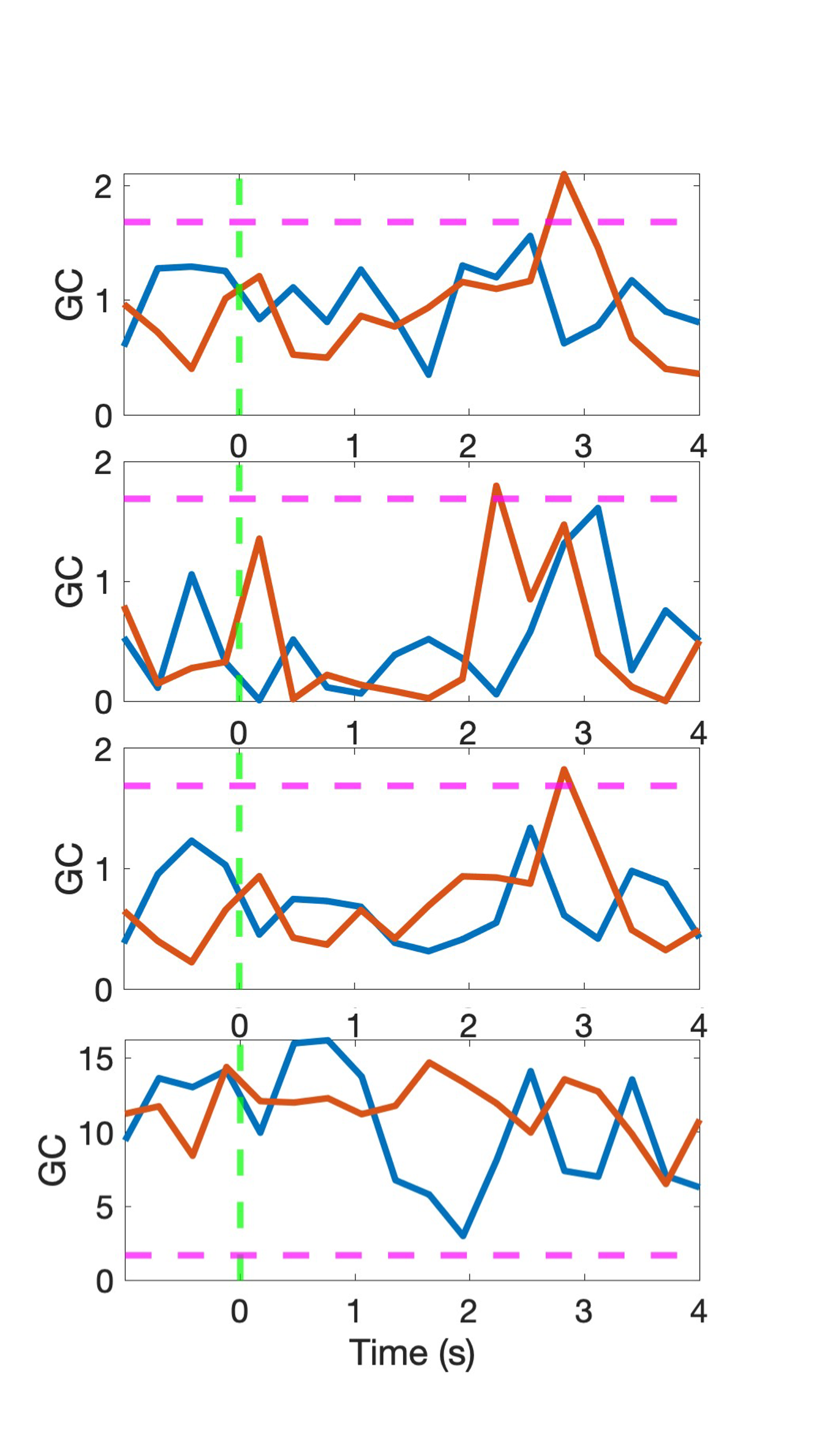}}
\subfigure[\textit{\small Subject K}]{\includegraphics[trim={0.2cm 1cm 1.5cm 1cm},clip,height = 4.6cm, width = 2.1cm]{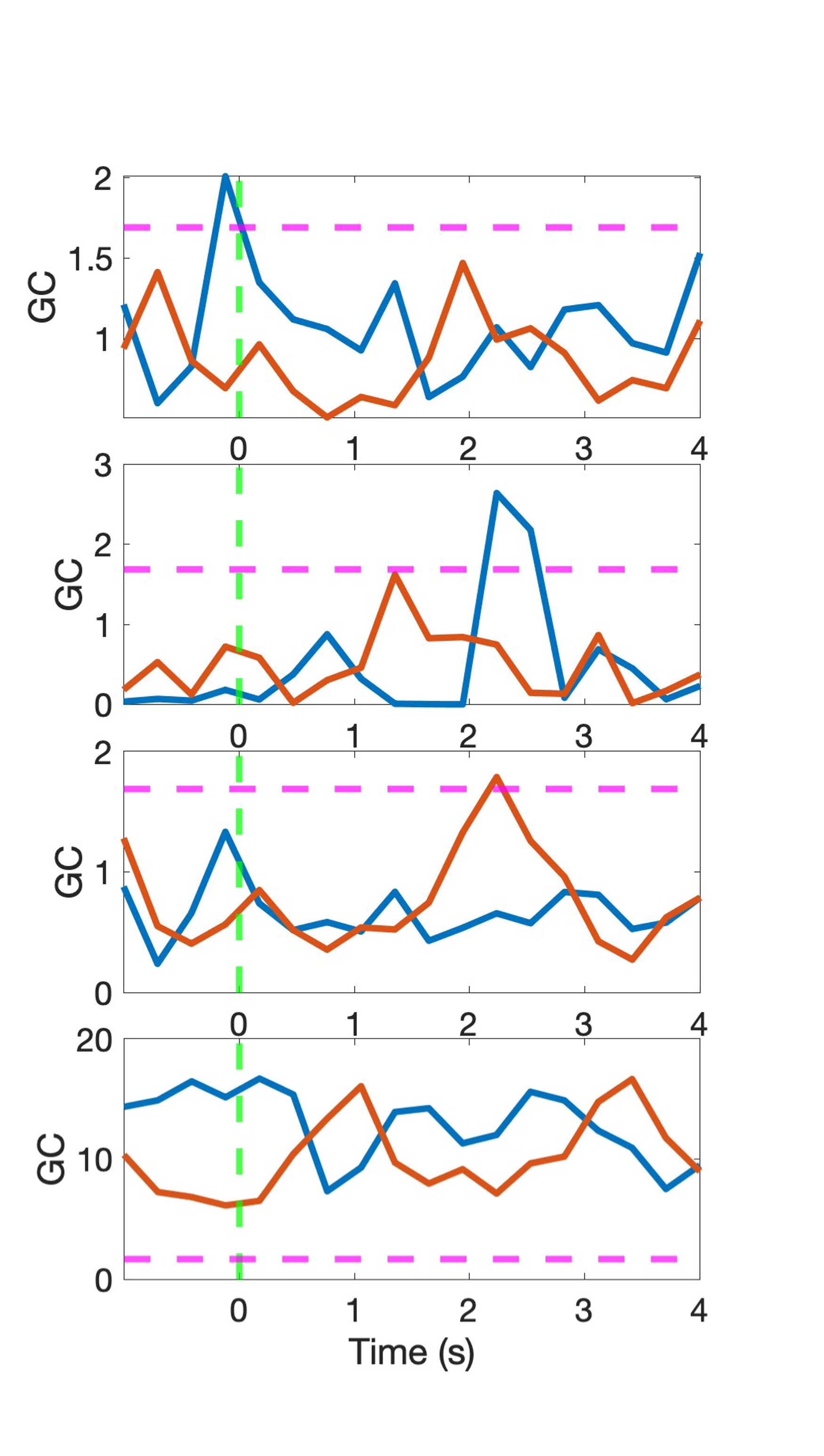}}
\subfigure[\textit{\small Subject L}]{\includegraphics[trim={0.2cm 1cm 1.5cm 1cm},clip,height = 4.6cm, width = 2.1cm]{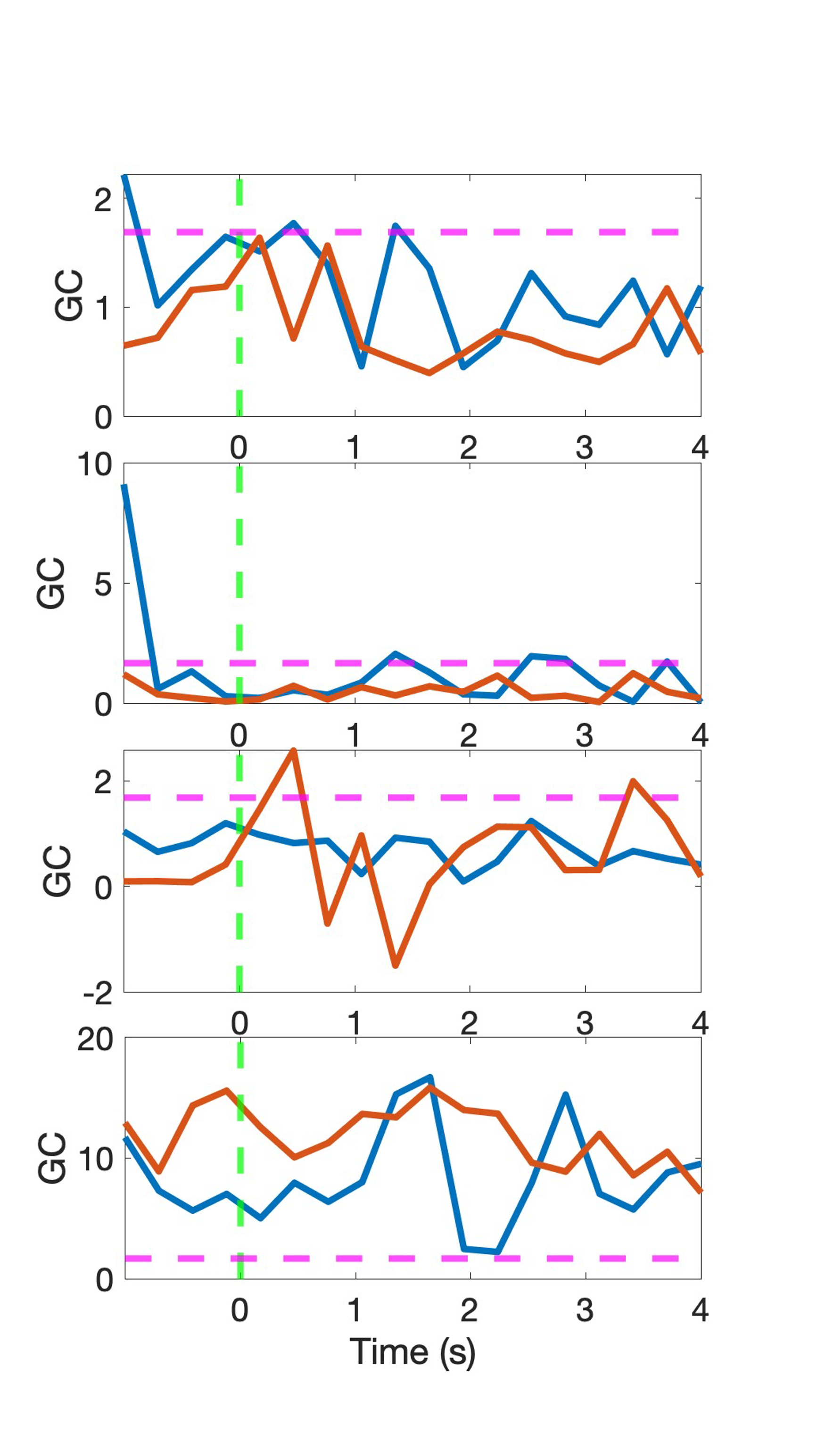}}
\subfigure[\textit{\small Subject M}]{\includegraphics[trim={0.2cm 1cm 1.5cm 1cm},clip,height = 4.6cm, width = 2.1cm]{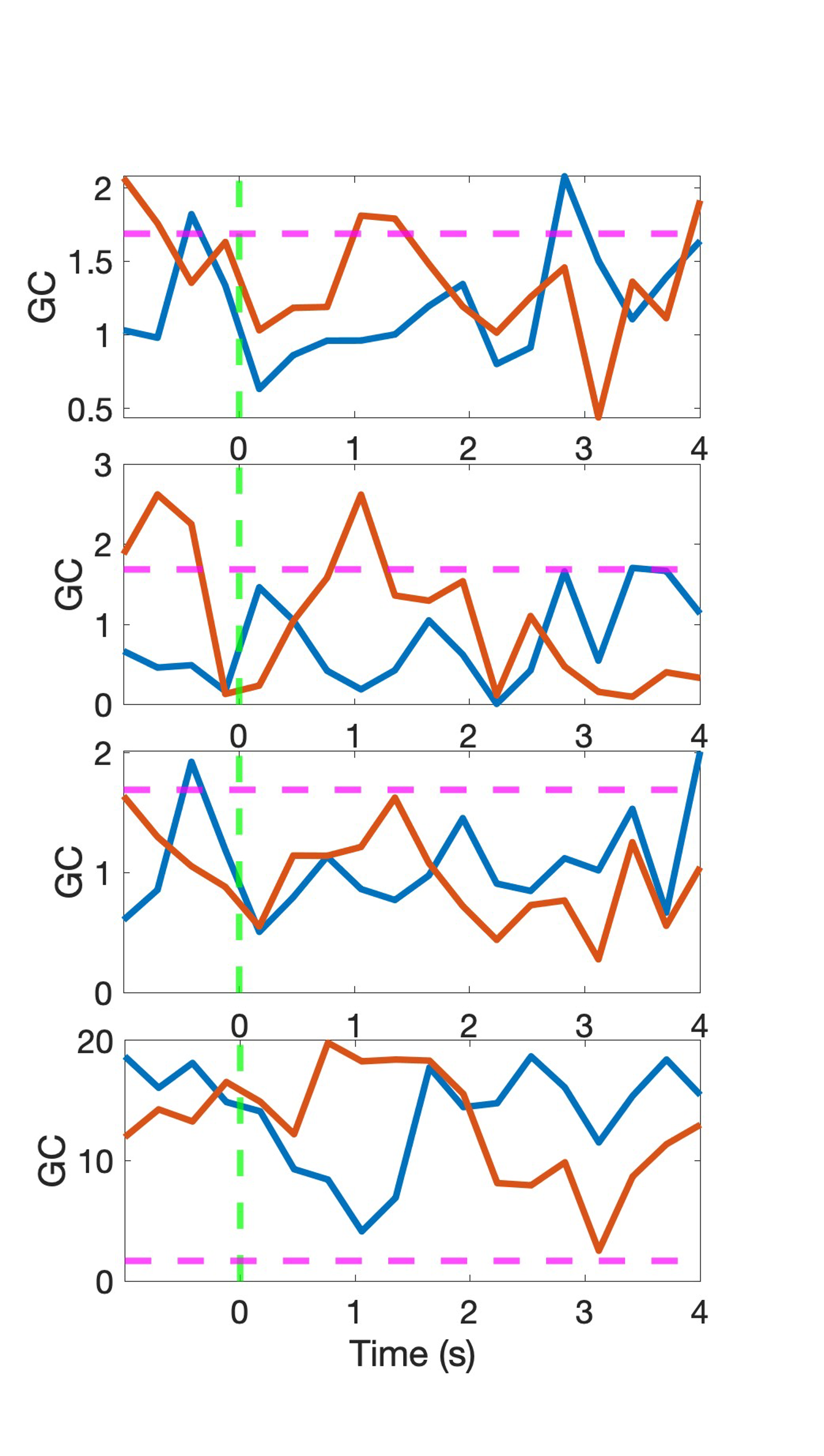}}
\subfigure[\textit{\small Subject N}]{\includegraphics[trim={0.2cm 1cm 1.5cm 1cm},clip,height = 4.6cm, width = 2.1cm]{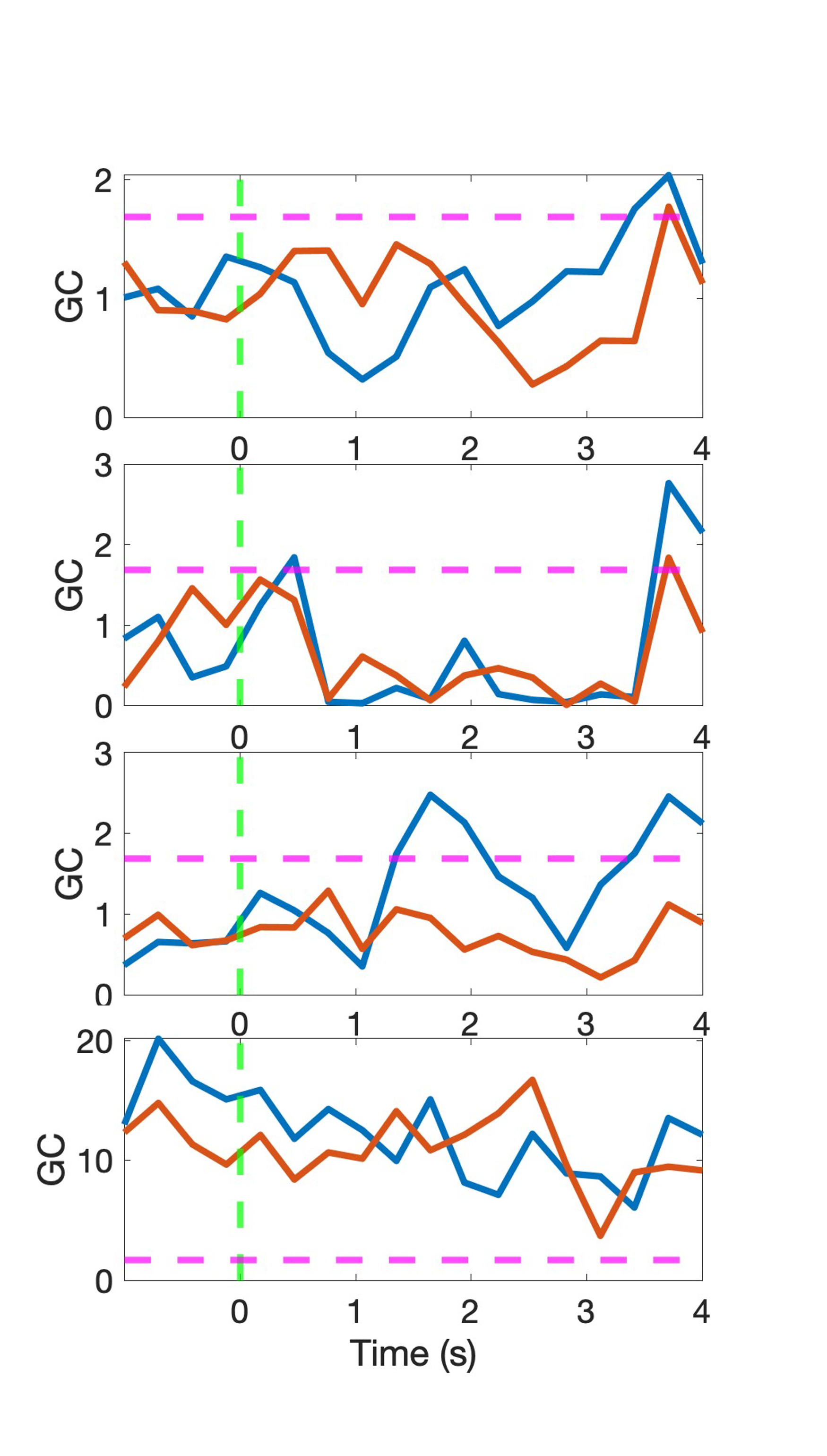}}
\subfigure[\textit{\small Subject Q}]{\includegraphics[trim={0.2cm 1cm 1.5cm 1cm},clip,height = 4.6cm, width = 2.1cm]{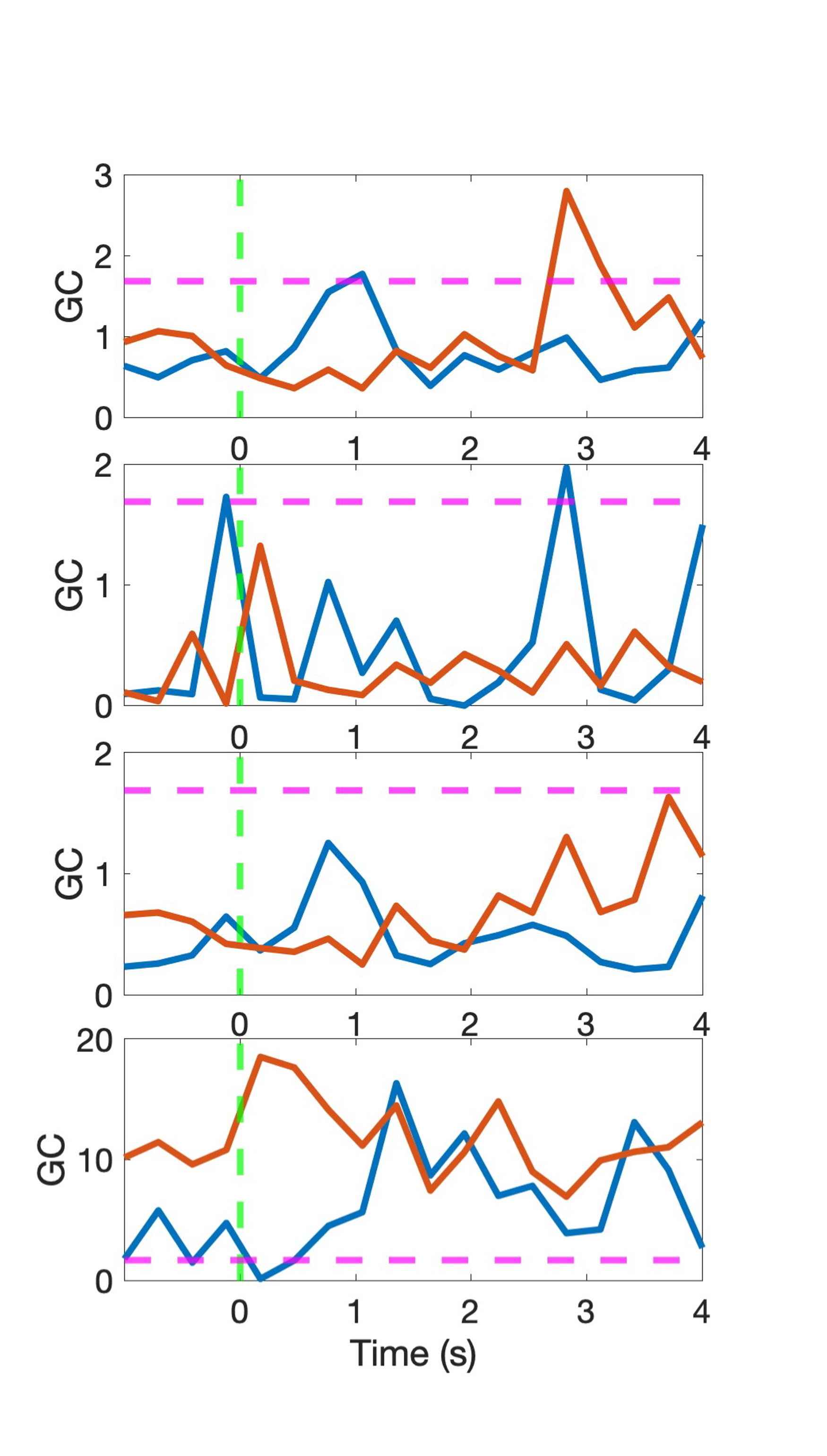}}
\centering
\subfigure{\includegraphics[height = 0.7cm, width = 9cm]{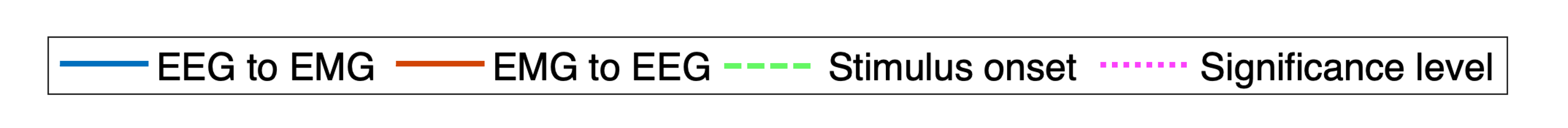}}
\caption{\textit{GC for eight healthy subjects. \textbf{First row:} Traditional GC \cite{lutkepohl2005new}, \textbf{Second row:} VAR (BIC+Wiener filtering \cite{somers2018generic}), \textbf{Third row:} SS-ADMM, \textbf{Fourth row:} SSD-ADMM}}
\label{fig4}
\end{figure}

{In order to  evaluate how cortex-muscle interactions might vary over the 5 second data epoch in relation to the stimulus, we employed a half-overlapping sliding window of 500ms duration to estimate causality over successive time intervals, as is typically performed in motor neuroscience studies \cite{lutkepohl2005new}.  The same approach was applied with each method being evaluated as indicated in \cite{guo2021unravelling}, \cite{guo2024subband}. Figure \ref{fig4} shows the resulting Granger causality over the 5 second data epoch, as detected using our proposed algorithms SS-ADMM and SSD-ADMM in comparison with the traditional Granger causality \cite{lutkepohl2005new}. As another comparative approach, we utilized a Wiener filter  \cite{somers2018generic} as a pre-processing step to effectively denoise the signals, after which we fitted a VAR model and employed the Bayesian Information Criterion (BIC) for model order identification. These results are also compared in Figure \ref{fig4} .   

}

Granger causality computed from all four algorithms across the 5 second epoch is shown for eight subjects.  An F-test applied with 95\% significance level tests the null hypothesis that the first time series does not  Granger-cause the second. In Figure \ref{fig4}, the vertical and horizontal lines represent stimulus onset and critical values of the F-test, respectively. In this active task we expect causality between brain and muscles, so the observed pre-stimulus causality in some subjects in Figure \ref{fig4}, is not unanticipated. 
The results in Figure 
\ref{fig4} demonstrate that Granger causality detection has been significantly improved by modelling the measurement noise.
The causality from sEMG to EEG signals has been discovered in many subjects using the proposed method SSD-ADMM that has not been discovered earlier using the other approaches. 

\section{Discussion} 

In this study, we explored the challenges associated with detecting cortico-muscular interactions during muscle activities in diverse subjects. These challenges arise from various factors, including measurement noise introduced by environmental interference, sensor imperfections, and quantization errors. While measurement noise is inevitable to some extent, we can take steps to minimize it at the time of recording through techniques such as shielding, filtering, calibration, and advanced signal processing methods. Nevertheless, some measurement noise will remain.  Additionally, the recordings will contain physiological \textit{noise}, which cannot be removed at the time of recording. In this paper we focused on a signal processing method to account for both the measurement noise and the physiological noise (or \textit{excitation noise}) in the data. Notably, for healthy subjects, modeling the noise revealed cortico-muscular interactions that were previously suppressed. In most of the subjects, no causality was detected above the critical level when using the standard "traditional" GC, but significant causal interaction was detected both pre- and post-stimulus after accounting for both the measurement and physiological noise, aligning with expectations for a healthy subject.

It is important to note that Granger causality serves as a tool for identifying statistical associations, and the observed causality should not be considered true causality in a causal inference sense. Establishing true causality typically demands experimental design, control groups, and a clear theoretical understanding of cause-and-effect relationships. While Granger causality does not imply true causality, it can assist in identifying statistical associations and temporal patterns in data, which may be of interest for further investigation.

Granger causality analysis has limitations, including assumptions of linearity and stationarity. Existing literature employs alternative metrics such as Mutual Information \cite{jin2010linear}, and Maximal Information Coefficient \cite{liang2020time}, considering the frequency spectrum of the recorded signals to capture frequency-specific non-linear interactions. While these methods offer a non-parametric way to extract associations, in some cases they lack directional information, are sensitive to noise, and may overestimate dependencies \cite{papana2009evaluation}. To measure the directed flow of information between time series, one of the widely used metrics is Transfer Entropy \cite{lindner2011trentool}, \cite{guo2021multiscale}. Empirical mode decomposition, often used in conjunction with Transfer Entropy \cite{cheng2019functional}, \cite{liu2020multiscale}, captures both non-linear and non-stationary interactions. However, it is computationally intensive and requires careful parameter tuning. The choice of appropriate methodology should consider data-specific characteristics and research objectives. For discerning non-linear or frequency-specific interactions, methods like cross-frequency coherence \cite{guo2023structured}, mutual information, or transfer entropy with empirical mode decomposition may be preferable. Conversely, if simplicity and computational efficiency are prioritized, Granger causality analysis, relying on autoregressive models, could be a preferred choice.

{
A further consideration is that the classical Granger causality approach, which uses autoregressive models for full and restricted forms, is subject to biases induced by the finite order of the restricted model\cite{stokes2017study, faes2017interpretability, antonacci2020information, antonacci2021estimation}. However, it remains a practical tool in many contexts. In the current study where the focus is on short-term causal relationships, autoregressive models can still provide reliable results without the complexity of more advanced techniques. Their simplicity and speed may outweigh potential biases, especially in situations where computational efficiency is a priority. As a direction for future research, we intend to investigate state-space representations of physiological data and compare with the current approach.}

In a comprehensive study on Granger causality \cite{shojaie2022granger}, the authors discuss limitations encountered in reliably discovering causal interactions in real systems. This research specifically addresses three of these limitations: known lag, stationarity, and perfect observation. As a future research direction, we aim to also mitigate the additional limitations while advancing existing methodologies to improve the reliability of the  detection of cortico-muscular interactions not only in healthy subjects but also in individuals with neurological disorders.

\section{Conclusion}

To conclude, we propose a novel way to find a proximal operator for regularization based on the stationarity condition augmented with the model identification in a convex optimization program. The resulting program can be solved efficiently using ADMM to guarantee a global optimal solution. The model orders for different  sets of parameters are identified up to an upper bound. We further refine the proposed approach to denoise autoregressive signals by assuming a wavelet sparsity assumption on the excitation noise term. We demonstrate that the proposed method effectively disentangles the measurement noise and the excitation noise terms consequently improving Granger causality detection in physiological data. Experimental results demonstrate the effectiveness of our approach for real-world data in detecting causal interactions.

\bibliographystyle{IEEEtran}  

\bibliography{maintex}  

\begin{thebibliography}{10}
\providecommand{\url}[1]{#1}
\csname url@samestyle\endcsname
\providecommand{\newblock}{\relax}
\providecommand{\bibinfo}[2]{#2}
\providecommand{\BIBentrySTDinterwordspacing}{\spaceskip=0pt\relax}
\providecommand{\BIBentryALTinterwordstretchfactor}{4}
\providecommand{\BIBentryALTinterwordspacing}{\spaceskip=\fontdimen2\font plus
\BIBentryALTinterwordstretchfactor\fontdimen3\font minus \fontdimen4\font\relax}
\providecommand{\BIBforeignlanguage}[2]{{%
\expandafter\ifx\csname l@#1\endcsname\relax
\typeout{** WARNING: IEEEtran.bst: No hyphenation pattern has been}%
\typeout{** loaded for the language `#1'. Using the pattern for}%
\typeout{** the default language instead.}%
\else
\language=\csname l@#1\endcsname
\fi
#2}}
\providecommand{\BIBdecl}{\relax}
\BIBdecl

\bibitem{delezie2018endocrine}
D.~Julien \emph{et~al.}, ``Endocrine crosstalk between skeletal muscle and the brain,'' \emph{Frontiers in neurology}, vol.~9, p. 698, 2018.

\bibitem{chowdhury2019eeg}
A.~Chowdhury \emph{et~al.}, ``An eeg-emg correlation-based brain-computer interface for hand orthosis supported neuro-rehabilitation,'' \emph{Journal of neuroscience methods}, vol. 312, pp. 1--11, 2019.

\bibitem{fuchs2009embodied}
T.~Fuchs, ``Embodied cognitive neuroscience and its consequences for psychiatry,'' \emph{Poiesis \& Praxis}, vol.~6, pp. 219--233, 2009.

\bibitem{chiel2009brain}
H.~J. Chiel \emph{et~al.}, ``The brain in its body: motor control and sensing in a biomechanical context,'' \emph{Journal of Neuroscience}, vol.~29, no.~41, pp. 12\,807--12\,814, 2009.

\bibitem{qiao2021survey}
H.~Qiao \emph{et~al.}, ``A survey of brain-inspired intelligent robots: Integration of vision, decision, motion control, and musculoskeletal systems,'' \emph{IEEE Transactions on Cybernetics}, vol.~52, no.~10, pp. 11\,267--11\,280, 2021.

\bibitem{mcclelland2012modulation}
V.~M. McClelland \emph{et~al.}, ``Modulation of corticomuscular coherence by peripheral stimuli,'' \emph{Experimental brain research}, vol. 219, pp. 275--292, 2012.

\bibitem{du2021dictionary}
S.~Du \emph{et~al.}, ``Dictionary learning strategies for cortico-muscular coherence detection and estimation,'' in \emph{2021 43rd Annual International Conference of the IEEE Engineering in Medicine \& Biology Society (EMBC)}.\hskip 1em plus 0.5em minus 0.4em\relax IEEE, 2021, pp. 240--244.

\bibitem{mcclelland2020abnormal}
V.~M. McClelland \emph{et~al.}, ``Abnormal patterns of corticomuscular and intermuscular coherence in childhood dystonia,'' \emph{Clinical Neurophysiology}, vol. 131, no.~4, pp. 967--977, 2020.

\bibitem{witham2011contributions}
C.~L. Witham \emph{et~al.}, ``Contributions of descending and ascending pathways to corticomuscular coherence in humans,'' \emph{The Journal of physiology}, vol. 589, no.~15, pp. 3789--3800, 2011.

\bibitem{zandvoort2019human}
C.~S. Zandvoort \emph{et~al.}, ``The human sensorimotor cortex fosters muscle synergies through cortico-synergy coherence,'' \emph{Neuroimage}, vol. 199, pp. 30--37, 2019.

\bibitem{hsiao1982autoregressive}
C.~Hsiao, ``Autoregressive modeling and causal ordering of economic variables,'' \emph{Journal of Economic Dynamics and Control}, vol.~4, pp. 243--259, 1982.

\bibitem{lutkepohl2005new}
H.~L{\"u}tkepohl, \emph{New introduction to multiple time series analysis}.\hskip 1em plus 0.5em minus 0.4em\relax Springer Science \& Business Media, 2007.

\bibitem{goebel2003investigating}
R.~Goebel \emph{et~al.}, ``Investigating directed cortical interactions in time-resolved fmri data using vector autoregressive modeling and {Granger} causality mapping,'' \emph{Magnetic resonance imaging}, vol.~21, no.~10, pp. 1251--1261, 2003.

\bibitem{ahmad2024smart}
H.~Ahmad, C.~Treude, M.~Wagner, and C.~Szabo, ``Smart hpa: A resource-efficient horizontal pod auto-scaler for microservice architectures,'' \emph{arXiv preprint arXiv:2403.07909}, 2024.

\bibitem{ahmad2024towards}
------, ``Towards resource-efficient reactive and proactive auto-scaling for microservice architectures,'' \emph{Available at SSRN 4918202}, 2024.

\bibitem{goel2024machine}
D.~Goel, H.~Ahmad, A.~K. Jain, and N.~K. Goel, ``Machine learning driven smishing detection framework for mobile security,'' \emph{arXiv preprint arXiv:2412.09641}, 2024.

\bibitem{ahmad2023review}
H.~Ahmad, I.~Dharmadasa, F.~Ullah, and M.~A. Babar, ``A review on c3i systems’ security: Vulnerabilities, attacks, and countermeasures,'' \emph{ACM Computing Surveys}, vol.~55, no.~9, pp. 1--38, 2023.

\bibitem{ahmad2024survey}
H.~Ahmad, F.~Ullah, and R.~Jafri, ``A survey on immersive cyber situational awareness systems,'' \emph{arXiv preprint arXiv:2408.07456}, 2024.

\bibitem{ahmad2025future}
H.~Ahmad and D.~Goel, ``The future of ai: Exploring the potential of large concept models,'' \emph{arXiv preprint arXiv:2501.05487}, 2025.

\bibitem{chopra2024chatnvd}
S.~Chopra, H.~Ahmad, D.~Goel, and C.~Szabo, ``Chatnvd: Advancing cybersecurity vulnerability assessment with large language models,'' \emph{arXiv preprint arXiv:2412.04756}, 2024.

\bibitem{haque2022think}
M.~U. Haque, I.~Dharmadasa, Z.~T. Sworna, R.~N. Rajapakse, and H.~Ahmad, ``" i think this is the most disruptive technology": Exploring sentiments of chatgpt early adopters using twitter data,'' \emph{arXiv preprint arXiv:2212.05856}, 2022.

\bibitem{abdulsatar2024towards}
M.~Abdulsatar, H.~Ahmad, D.~Goel, and F.~Ullah, ``Towards deep learning enabled cybersecurity risk assessment for microservice architectures,'' \emph{arXiv preprint arXiv:2403.15169}, 2024.

\bibitem{jayalath2024microservice}
R.~K. Jayalath, H.~Ahmad, D.~Goel, M.~S. Syed, and F.~Ullah, ``Microservice vulnerability analysis: A literature review with empirical insights,'' \emph{IEEE Access}, 2024.

\bibitem{granger2001investigating}
C.~Granger, ``Investigating causal relations by econometric models and cross-spectral methods,'' in \emph{Essays in econometrics: collected papers of Clive WJ Granger}, 2001, pp. 31--47.

\bibitem{shojaie2022granger}
A.~Shojaie and E.~B. Fox, ``Granger causality: A review and recent advances,'' \emph{Annual Review of Statistics and Its Application}, vol.~9, pp. 289--319, 2022.

\bibitem{florin2016parkinson}
E.~Florin \emph{et~al.}, ``Parkinson subtype-specific {Granger}-causal coupling and coherence frequency in the subthalamic area,'' \emph{Neuroscience}, vol. 332, pp. 170--180, 2016.

\bibitem{akaike1969fitting}
H.~Akaike, ``Fitting autoregressive models for prediction,'' \emph{Annals of the institute of Statistical Mathematics}, vol.~21, no.~1, pp. 243--247, 1969.

\bibitem{schwarz1978estimating}
G.~Schwarz, ``Estimating the dimension of a model,'' \emph{The annals of statistics}, pp. 461--464, 1978.

\bibitem{shmueli2010explain}
G.~Shmueli, ``To explain or to predict?'' 2010.

\bibitem{cruz2006good}
N.~Cruz-Ram{\'\i}rez \emph{et~al.}, ``How good are the bayesian information criterion and the minimum description length principle for model selection? a bayesian network analysis,'' in \emph{Mexican International Conference on Artificial Intelligence}.\hskip 1em plus 0.5em minus 0.4em\relax Springer, 2006, pp. 494--504.

\bibitem{liu2020fitting}
Y.~Liu and S.~D. Tajbakhsh, ``Fitting arma time series models without identification: A proximal approach,'' \emph{arXiv preprint arXiv:2002.06777}, 2020.

\bibitem{abbas2024dlgc}
F.~Abbas, V.~McClelland, Z.~Cvetkovic, and W.~Dai, ``Dlgc: Dictionary learning based granger causal discovery for cortico-muscular coupling,'' in \emph{2024 32nd European Signal Processing Conference (EUSIPCO)}.\hskip 1em plus 0.5em minus 0.4em\relax IEEE, 2024, pp. 1746--1750.

\bibitem{songsiri2013sparse}
J.~Songsiri, ``Sparse autoregressive model estimation for learning {Granger} causality in time series,'' in \emph{2013 IEEE International Conference on Acoustics, Speech and Signal Processing}.\hskip 1em plus 0.5em minus 0.4em\relax IEEE, 2013, pp. 3198--3202.

\bibitem{watson1986univariate}
M.~W. Watson, ``Univariate detrending methods with stochastic trends,'' \emph{Journal of monetary economics}, vol.~18, no.~1, pp. 49--75, 1986.

\bibitem{graupe1980convergence}
D.~Graupe, ``On convergence of least-squares identifiers of autoregressive models having stable and unstable roots,'' \emph{IEEE Transactions on Automatic Control}, vol.~25, no.~5, pp. 999--1002, 1980.

\bibitem{patriota2010vector}
A.~G. Patriota \emph{et~al.}, ``Vector autoregressive models with measurement errors for testing {Granger} causality,'' \emph{Statistical Methodology}, vol.~7, no.~4, pp. 478--497, 2010.

\bibitem{alyasseri2021eeg}
Z.~A.~A. Alyasseri \emph{et~al.}, ``{EEG} signal denoising using hybridizing method between wavelet transform with genetic algorithm,'' in \emph{Proceedings of the 11th National Technical Seminar on Unmanned System Technology 2019: NUSYS'19}.\hskip 1em plus 0.5em minus 0.4em\relax Springer, 2021, pp. 449--469.

\bibitem{abbas2024robust}
F.~Abbas and H.~Ahmad, ``Robust partial least squares using low rank and sparse decomposition,'' \emph{arXiv preprint arXiv:2407.06936}, 2024.

\bibitem{guo2023structured}
Z.~Guo \emph{et~al.}, ``Structured {Errors-In-Variables} modelling for cortico-muscular coherence enhancement,'' in \emph{ICASSP 2023}.\hskip 1em plus 0.5em minus 0.4em\relax IEEE, 2023.

\bibitem{soderstrom2003errors}
T.~S{\"o}derstr{\"o}m, ``Why are {Errors-In-Variables} problems often tricky?'' in \emph{2003 European Control Conference (ECC)}.\hskip 1em plus 0.5em minus 0.4em\relax IEEE, 2003, pp. 802--807.

\bibitem{christmas2010robust}
J.~Christmas and R.~Everson, ``Robust autoregression: Student-t innovations using variational bayes,'' \emph{IEEE Transactions on Signal Processing}, vol.~59, no.~1, pp. 48--57, 2010.

\bibitem{stokes2017study}
P.~A. Stokes and P.~L. Purdon, ``A study of problems encountered in granger causality analysis from a neuroscience perspective,'' \emph{Proceedings of the national academy of sciences}, vol. 114, no.~34, pp. E7063--E7072, 2017.

\bibitem{antonacci2020information}
Y.~Antonacci \emph{et~al.}, ``Information transfer in linear multivariate processes assessed through penalized regression techniques: validation and application to physiological networks,'' \emph{Entropy}, vol.~22, no.~7, p. 732, 2020.

\bibitem{faes2017interpretability}
L.~Faes \emph{et~al.}, ``On the interpretability and computational reliability of frequency-domain granger causality,'' \emph{F1000Research}, vol.~6, 2017.

\bibitem{antonacci2021estimation}
Y.~Antonacci \emph{et~al.}, ``Estimation of granger causality through artificial neural networks: applications to physiological systems and chaotic electronic oscillators,'' \emph{PeerJ Computer Science}, vol.~7, p. e429, 2021.

\bibitem{abbas2023ss}
F.~Abbas \emph{et~al.}, ``Ss-admm: Stationary and sparse granger causal discovery for cortico-muscular coupling,'' in \emph{ICASSP 2023-2023 IEEE International Conference on Acoustics, Speech and Signal Processing (ICASSP)}.\hskip 1em plus 0.5em minus 0.4em\relax IEEE, 2023, pp. 1--5.

\bibitem{jacob2009group}
L.~Jacob \emph{et~al.}, ``Group lasso with overlap and graph lasso,'' in \emph{Proceedings of the 26th annual international conference on machine learning}, 2009, pp. 433--440.

\bibitem{wang2013projection}
W.~Wang and M.~A. Carreira-Perpin{\'a}n, ``Projection onto the probability simplex: An efficient algorithm with a simple proof, and an application,'' \emph{arXiv preprint arXiv:1309.1541}, 2013.

\bibitem{nalatore2007mitigating}
H.~Nalatore \emph{et~al.}, ``Mitigating the effects of measurement noise on granger causality,'' \emph{Physical Review E}, vol.~75, no.~3, p. 031123, 2007.

\bibitem{abbas2024infr}
F.~Abbas, V.~McClelland, Z.~Cvetkovic, and W.~Dai, ``Infr-gc: Interpretable feature representations for granger causality in cortico-muscular interactions,'' in \emph{IEEE International Conference on Acoustics, Speech and Signal Processing, ICASSP 2025}, 2024.

\bibitem{ccayir2021maximum}
{\"O}.~{\c{C}}ay{\i}r and {\c{C}}.~Candan, ``Maximum likelihood autoregressive model parameter estimation with noise corrupted independent snapshots,'' \emph{Signal Processing}, vol. 186, p. 108118, 2021.

\bibitem{park2019measurement}
S.~Park \emph{et~al.}, ``Measurement noise recommendation for efficient kalman filtering over a large amount of sensor data,'' \emph{Sensors}, vol.~19, no.~5, p. 1168, 2019.

\bibitem{averkamp2003wavelet}
R.~Averkamp and C.~Houdr{\'e}, ``Wavelet thresholding for non-necessarily gaussian noise: idealism,'' \emph{The Annals of Statistics}, vol.~31, no.~1, pp. 110--151, 2003.

\bibitem{antoniadis2002wavelet}
A.~Antoniadis \emph{et~al.}, ``Wavelet thresholding for some classes of non--gaussian noise,'' \emph{Statistica neerlandica}, vol.~56, no.~4, pp. 434--453, 2002.

\bibitem{box2015time}
G.~E. Box \emph{et~al.}, \emph{Time series analysis: forecasting and control}.\hskip 1em plus 0.5em minus 0.4em\relax John Wiley \& Sons, 2015.

\bibitem{shekar2019grid}
B.~Shekar and G.~Dagnew, ``Grid search-based hyperparameter tuning and classification of microarray cancer data,'' in \emph{2019 second international conference on advanced computational and communication paradigms (ICACCP)}.\hskip 1em plus 0.5em minus 0.4em\relax IEEE, 2019, pp. 1--8.

\bibitem{vogel1996iterative}
C.~R. Vogel and M.~E. Oman, ``Iterative methods for total variation denoising,'' \emph{SIAM Journal on Scientific Computing}, vol.~17, no.~1, pp. 227--238, 1996.

\bibitem{guo2021unravelling}
Z.~Guo \emph{et~al.}, ``Unravelling causal relationships between cortex and muscle with {Errors-In-Variables} models,'' in \emph{2021 43rd Annual International Conference of the IEEE Engineering in Medicine \& Biology Society (EMBC)}.\hskip 1em plus 0.5em minus 0.4em\relax IEEE, 2021, pp. 967--970.

\bibitem{guo2024subband}
------, ``Subband independent component analysis for coherence enhancement,'' \emph{IEEE Transactions on Biomedical Engineering}, 2024.

\bibitem{somers2018generic}
B.~Somers \emph{et~al.}, ``A generic eeg artifact removal algorithm based on the multi-channel wiener filter,'' \emph{Journal of neural engineering}, vol.~15, no.~3, p. 036007, 2018.

\bibitem{jin2010linear}
S.-H. Jin \emph{et~al.}, ``Linear and nonlinear information flow based on time-delayed mutual information method and its application to corticomuscular interaction,'' \emph{Clinical Neurophysiology}, vol. 121, no.~3, pp. 392--401, 2010.

\bibitem{liang2020time}
T.~Liang \emph{et~al.}, ``Time-frequency maximal information coefficient method and its application to functional corticomuscular coupling,'' \emph{IEEE Transactions on Neural Systems and Rehabilitation Engineering}, vol.~28, no.~11, pp. 2515--2524, 2020.

\bibitem{papana2009evaluation}
A.~Papana and D.~Kugiumtzis, ``Evaluation of mutual information estimators for time series,'' \emph{International Journal of Bifurcation and Chaos}, vol.~19, no.~12, pp. 4197--4215, 2009.

\bibitem{lindner2011trentool}
M.~Lindner \emph{et~al.}, ``Trentool: A matlab open source toolbox to analyse information flow in time series data with transfer entropy,'' \emph{BMC neuroscience}, vol.~12, pp. 1--22, 2011.

\bibitem{guo2021multiscale}
Z.~Guo \emph{et~al.}, ``Multiscale wavelet transfer entropy with application to corticomuscular coupling analysis,'' \emph{IEEE Transactions on Biomedical Engineering}, vol.~69, no.~2, pp. 771--782, 2021.

\bibitem{cheng2019functional}
S.~Cheng \emph{et~al.}, ``Functional corticomuscular coupling based on bivariate empirical mode decomposition-multiscale transfer entropy,'' in \emph{2019 IEEE International Conference on Computational Intelligence and Virtual Environments for Measurement Systems and Applications (CIVEMSA)}.\hskip 1em plus 0.5em minus 0.4em\relax IEEE, 2019, pp. 1--4.

\bibitem{liu2020multiscale}
J.~Liu \emph{et~al.}, ``Multiscale transfer spectral entropy for quantifying corticomuscular interaction,'' \emph{IEEE Journal of Biomedical and Health Informatics}, vol.~25, no.~6, pp. 2281--2292, 2020.

\end{thebibliography}

\end{document}